\documentclass[preprint,article,accept,moreauthors,pdftex]{Definitions/mdpi} 
\usepackage{algorithm}      
\usepackage{algorithmic}    
\usepackage{soul}
\soulregister\cite7 
\soulregister\citep7 
\soulregister\citet7 
\soulregister\ref7 
\soulregister\pageref7 
\soulregister\textit7 

\usepackage{tabularx}  
\usepackage{makecell}   
\usepackage{enumitem}   
\usepackage{xcolor}
\sethlcolor{yellow}
\usepackage{array} 
\newcolumntype{P}[1]{>{\raggedright\arraybackslash}p{#1}}
\firstpage{1} 
\pubvolume{15}
\issuenum{13}
\articlenumber{7227}
\pubyear{2025}
\copyrightyear{2025}
\externaleditor{Douglas O'Shaughnessy} 
\datereceived{19 May 2025} 
\daterevised{18 June 2025} 
\dateaccepted{24 June 2025} 
\datepublished{26 June 2025} 
\hreflink{https://\linebreak doi.org/10.3390/app15137227} 

\Title{Large-Language-Model-Enabled 
	Text Semantic Communication~Systems}

\TitleCitation{Large-\linebreak Language-Model-Enabled Text Semantic Communication Systems}
\Author {Zhenyi Wang 
$^{1}$
\orcidA{}, Li Zou $^{1}$, Shengyun Wei $^{2}$\orcidB{}, Kai Li $^{1,}$*\orcidD{}, Feifan Liao $^{1}$, Haibo Mi $^{1}$ and Rongxuan Lai $^{1,}$*\orcidC{}}

\AuthorNames{Zhenyi Wang, Li Zou, Shengyun Wei, Kai Li, Feifan Liao, Haibo Mi and Rongxuan Lai}

\isAPAStyle{%
       \AuthorCitation{Wang, Z.;  Zou, L.; Wei, S.; Li, K.; Liao, F.; Mi H and Lai, R.}
         }{%
        \isChicagoStyle{%
        \AuthorCitation{Lastname, Firstname, Firstname Lastname, and Firstname Lastname.}
        }{
        \AuthorCitation{Wang, Z.; 
 Zou, L.; Wei, S.; Li, K.; Liao, F.; Mi, H.; Lai, R.}
        }
}

\address{%
$^{1}$ \quad  College of Information and Communication, National University of Defense Technology, \linebreak Wuhan 430000, 
China; wangzhenyi@nudt.edu.cn (Z.W.); zouli17@nudt.edu.cn (L.Z.);  
\linebreak  liaofeifan17@nudt.edu.cn (F.L.); haibo\_mihb@126.com (H.M.)\\ 
$^{2}$ \quad School of Electrical Engineering, Naval University of Engineering,  
Wuhan 430000, 
 China; shengyunwei@nudt.edu.cn}

\corres{Correspondence: lokiavera1991@163.com (K.L.); lairongxuan123@163.com (R.L.) 
}

\abstract{Large language models (LLMs) have recently demonstrated state-of-the-art performance in various natural language processing (NLP) tasks, achieving near-human levels in multiple language understanding challenges and aligning closely with the core principles of semantic communication  
Inspired by LLMs' advancements in semantic processing, we propose LLM-SC, an innovative LLM-enabled semantic communication system framework which applies LLMs directly to the physical layer coding and decoding for the first time. By analyzing the relationship between the training process of LLMs and the optimization objectives of semantic communication, we propose training a semantic encoder through LLMs' tokenizer training and establishing a semantic knowledge base via the LLMs' unsupervised pre-training process. This knowledge base facilitates the creation of optimal decoder by providing the prior probability of the transmitted language sequence. Based on this, we derive the optimal decoding criteria for the receiver and introduce beam search algorithm to further reduce complexity. Furthermore, we assert that existing LLMs can be employed directly for LLM-SC without extra re-training or fine-tuning. Simulation results reveal that LLM-SC outperforms conventional DeepSC at signal-to-noise ratios (SNRs) exceeding 3~dB, as it enables error-free transmissions of semantic information under high SNRs while DeepSC fails to do so. In addition to semantic-level performance, LLM-SC demonstrates compatibility with technical-level performance, achieving approximately an 8 dB coding gain for a bit error ratio (BER) of $10^{-3}$ without any channel coding while maintaining the same joint {source--channel} 
coding rate as traditional communication systems.}


\begin{document}

\section{Introduction 
}
Shannon and Weaver's seminal work categorizes communication systems into three hierarchical levels \citep{shannon1948mathematical, weaver1953recent}:  
\begin{itemize}
  \item Technical communication:
\textls[-55]{How accurately can the symbols of communication \mbox{be transmitted?}}
  \item Semantic communication:
 How precisely do the transmitted symbols convey the desired meaning? 
  \item Effective communication:
 How can the transmitted symbols be used to achieve the desired effect? 
\end{itemize}

Ever since Morse invented Morse code and the wired telegraph for communication in 1837, communication systems have consistently aimed to achieve reliable and efficient transmission of symbols without considering their meanings. Shannon's information theory, which quantifies information based on the probability of symbols, has  {provided a} theoretical  {basis} for all practical communication systems. As communication technology has advanced, the demands  {for} communication systems  {are steadily growing}. For instance, the development of 6G wireless communication systems significantly increased the demand for high-speed, low-latency communication to support various applications such as autonomous driving, the Internet of Things, and the Internet of Vehicles. This evolution necessitates the deployment of advanced coding techniques. Furthermore, the vast data transmission required  {by} emerging technologies, such as autonomous driving and the metaverse, are approaching the Shannon limit with traditional communication methodologies, thereby  {pressing for} new paradigms. Semantic communication, which emphasizes the precise transmission of semantic information, transcends conventional symbol-level source and channel coding. For instance, in the context of autonomous driving, it is often unnecessary to transmit raw sensor images to the control center; instead, the controller is primarily concerned with the semantics of the objects detected by the sensors.

Semantic communication systems have been studied for many years, with early discussions on the theory of semantic communication dating back to 1952 \citep{carnap1952outline}. Recent theoretical research has categorized semantic communication problem{s} into language exploitation and language design and provided mathematical models for both \citep{shao2024theory}. In practical engineering applications, DeepSC, the first semantic transmission system for text,  was proposed in \citep{xie2021deep}.  {Moreover}, DeepSC evaluates semantic communication systems  {by} using BLEU and sentence similarity metrics and introduces a transformer-based network for semantic encoding and decoding.  {Then}, several improved models have been developed  {on this basis} \citep{zhou2021semantic, xie2021task}. Furthermore, semantic communication is gradually being explored in  {the fields of} speech \citep{weng2021semantic}, images \citep{huang2021deep}, and video \citep{wang2022wireless}.

The rapid  {evolution} of artificial intelligence, particularly in NLP, continues to drive progress in semantic communication by providing new paradigms. LLMs have recently  {taken} {machines'} 
understanding of human language to new levels {\citep{chang2024survey, 11015742}.} 
 For instance, ChatGPT-3.5 
 can understand and respond to human language, well satisfying the needs of semantic communication.  {Therefore}, applying LLM technology to semantic communication systems is a natural progression, with some initial studies already exploring this integration \citep{guo2023semantic}. However, many issues in semantic communication remain unresolved, such as semantic representation and measurement, semantic error correction, and semantic knowledge bases~\citep{shi2018semantic}.

In this paper, we  {investigate} semantic communication systems by applying LLM technology to semantic encoding and decoding. A novel LLM-SC framework is proposed for text in wireless communication. The proposed framework utilizes LLM technology to model the semantic information of text and design a semantic communication system based on LLMs. We evaluate the framework through a series of simulations, demonstrating its effectiveness across multiple metrics. The main contributions of this paper are  {as follows}:
\begin{itemize}
  \item  {A novel LLM-SC framework for text in wireless communication featuring a dual-component structure: 
  (a) a semantic encoder 
utilizing LLM's tokenizer for joint source--channel coding, and (b) a semantic knowledge base 
built by LLM's unsupervised pre-training. This knowledge base provides prior probability distributions of language sequences, enabling optimal decoding at the receiver.} This is the first application of LLM to the physical-layer encoding and decoding in semantic communication. By using LLMs to  
probabilistically model transmitted language sequences, we achieve a communication paradigm that balances semantic-level and technical-level performance. We investigate the relationship between the training process of LLMs and the optimization goals of semantic communication, proposing the training of a semantic encoder using LLMs' tokenizer training and the construction of a semantic knowledge base using LLM unsupervised pre-training.
  \item The optimal detection for LLM-SC is derived, and beam search algorithm from the domain of NLP is introduced to reduce complexity. This algorithm strikes a balance in complexity between Viterbi decoding algorithm and greedy decoding algorithm, thereby ensuring  {both} high decoding efficiency and maintaining decoding reliability. Comparative simulations demonstrate that exploring the optimal paths with a beam size in the order of tens yields quite excellent decoding performance compared to traversing a vocabulary of tens of thousands. Moreover, when the beam size exceeds 20, the performance improvements by further increasing the beam size become~negligible.
   \item The semantic-level and technical-level performance of LLM-SC under AWGN and Rayleigh fading channels are evaluated, without requiring any additional re-training and fine-tuning on the existing LLM. Results show that LLM-SC outperforms the classical DeepSC in semantic-level metrics at an SNR above 3 dB and can achieve error-free semantic transmission at high SNRs, a contribution unattainable by DeepSC; moreover, in terms of technical-level metrics, LLM-SC exhibits superior coding efficiency compared to classical text compression algorithms and achieves optimal BER performance across the entire SNR curve at the same equivalent joint source--channel coding rate.
\end{itemize}

\section{Related Works}

 {To provide a structured overview of existing research, Table \ref{tab:review_summary} summarizes key contributions and limitations in three domains: (1) AI-enabled wireless communication, (2)~semantic representation in NLP, and (3) LLM-enabled semantic communication.}

\begin{table}[H]
  \caption{ {Summary of related works.} \label{tab:review_summary}}

\begin{adjustwidth}{-\extralength}{0cm}
\centering 
  \begin{tabularx}{\fulllength}{CCcC}
    \toprule
    {\textbf{Research} \textbf{Dimension}} & 
    {\textbf{Paradigm} \textbf{Characteristics}} & 
    {\textbf{Related} \textbf{Works}} & 
    {\textbf{Limitations and} \textbf{\mbox{Research Gaps}}} \\
    \midrule
    AI-enabled wireless communication 
 & 
    Neural network-based, end-to-end, data-based
    & 
    \cite{azzouz1996modulation,suetrong2024enhanced,jia2023lightweight,choi2018necst,wang2024graph,yuan2018deep,soltani2019deep,hekland2009shannon}

    & 
    Limited to low-level signal processing; no semantic integration; no physical \mbox{layer integration} \\
    \midrule
    Semantic representation 
 & 
    Embedding spaces, \mbox{contextual encoding,} knowledge~structuring
 &
 \citep{devlin2018bert,radford2018improving,mikolov2013efficient,pennington2014glove,berant2013semantic,fellbaum2010wordnet,joulin2016bag,alhammadi2024artificial}

 & 
Not designed for communication; isolated from PHY layer; static representations; no \mbox{channel adaptation}
 \\
\midrule
LLM-enabled  frameworks 
 & 
 Generative semantics,
knowledge grounding,
 cross-modal alignment,
 &
 \citep{guo2023semantic,jiang2023large,jiang2024large,shen2024large,valmeekam2023llmzip,zhao2024lamosc}
 & 
 High complexity; shallow \mbox{PHY integration;} 
 \mbox{computational overhead} \\
    \bottomrule
  \end{tabularx}
\end{adjustwidth}
\end{table}

\subsection{AI-Enabled Wireless Communication}
The burgeoning field of artificial intelligence (AI) has catalyzed a profound transformation in wireless communications. Unlike traditional end-to-end communication algorithms that rely on mathematical theories, AI leverages neural networks trained to model the input--output relationships of communication systems. By utilizing large datasets, these neural networks can learn the inherent patterns within communication systems. This approach emphasizes end-to-end optimization and data-driven methodologies, which significantly reduce design overhead. Historical efforts to integrate AI into wireless communications date back to 1992, when the authors of \citep{azzouz1996modulation} attempted to use neural networks for modulation recognition. Contemporary AI-based methods have now surpassed classical mathematical algorithms in modulation recognition \citep{suetrong2024enhanced}. For text data transmission, the authors of \citep{jia2023lightweight} introduced a unique deep and lightweight Transformer variant, DeLighT, which achieves end-to-end joint source--channel coding (JSCC). Subsequently, neural network methods have been successfully applied to various domains, including neural error correction \citep{choi2018necst}, channel modeling \citep{wang2024graph}, encoding and decoding \citep{yuan2018deep}, channel estimation and equalization \citep{soltani2019deep}, and bandwidth compression mappings \citep{hekland2009shannon}. These methods have made impressive performance improvements in wireless communications. A comprehensive review of neural network-based wireless communication technologies is provided in~\citep{alhammadi2024artificial}. By leveraging the capabilities of AI, wireless communication systems can achieve higher efficiency, adaptability, and performance, marking a significant advancement over traditional methods.

\subsection{Semantic Representation in Natural Language Processing}
Recent advancements in NLP, primarily driven by large pre-trained language models such as Bidirectional Encoder Representations from Transformers (BERT) \citep{devlin2018bert}, Generative Pre-trained Transformer (GPT) \citep{radford2018improving}, and their successors, have substantially enhanced the capability to process and generate human language. These models are {remarkably proficient} in understanding context, generating coherent text, and performing a variety of language tasks, establishing them as pivotal in semantic communication. Techniques such as Word2Vec \citep{mikolov2013efficient}, GloVe \citep{pennington2014glove}, and FastText \citep{joulin2016bag} transform words into continuous vector spaces, wherein semantically similar words are positioned closer together. These embeddings have revolutionized numerous NLP tasks by effectively capturing semantic meaning. However, traditional word embeddings lack the ability to capture syntactic information. In contrast, models like BERT and Embeddings from Language Models (ELMo) generate dynamic embeddings for words based on their context within sentences, thereby enhancing the capture of polysemy and contextual nuances. Semantic parsing, which involves converting natural language into formal representations such as logical forms or knowledge graphs, is crucial for tasks like question-answering and information extraction~\citep{berant2013semantic}. Ontologies like WordNet \citep{fellbaum2010wordnet} and knowledge graphs such as Google's Knowledge Graph~\citep{chah2018ok} organize information into interconnected entities and relationships, offering a rich semantic framework that underpins various NLP applications. Advanced architectures, notably Transformers, have dramatically bolstered models' abilities to understand and generate language. These architectures leverage large-scale pre-training and fine-tuning on specific tasks, achieving state-of-the-art performance across numerous benchmarks.

\subsection{LLM-Enabled Semantic Communication}
Recent breakthroughs in LLMs have significantly impacted semantic communication across various domains \citep{guo2023semantic}. For instance, Jiang {et al.} addressed challenges in semantic communication for image data through a framework incorporating a segment-anything model-based knowledge base, attention-based semantic integration, and adaptive compression techniques \citep{jiang2023large}. Similarly, ref. \citep{jiang2024large} propose a semantic communication framework (LAM-SC) tailored for image data, leveraging LLMs as the core knowledge base. These approaches harness LLMs' deep understanding of human knowledge to construct robust knowledge bases for diverse communication tasks. Shen {et al.} exploited the capabilities of LLMs in language understanding, planning, and code generation, combined with classical command strategies like task-oriented and communication-edge joint learning. They proposed an efficient multifunctional framework for coordinating edge AI models in executing tasks of edge intelligence \citep{shen2024large}. However, current LLM-based semantic communication systems primarily focus on higher-level tasks such as user intent understanding, posing challenges in their application to physical-layer encoding and decoding. Nevertheless, LLMzip has achieved compression ratios surpassing previously known estimates of the entropy bounds by employing LLMs for lossless text compression \citep{valmeekam2023llmzip}. Y. Zhao et al. {put forward} a semantic communication system driven by LLMs (LLMs) that uses multimodal features to reconstruct raw visual information, thereby improving transmission quality of images \citep{zhao2024lamosc}. This demonstrates exceptional source coding performance at the physical layer, notwithstanding complexity issues. However, the feasibility and benefits of utilizing LLMs for channel encoding/decoding and modulation/demodulation {remain largely~unexplored.} 

\section{System Model and Mechanism}

\subsection{Problem Description}
The primary objective of a communication system is to effectively and reliably transmit information-bearing messages to the receiver. Within such systems, the information sequence {is subjected to} 
source encoding, channel encoding, and modulation at the transmitter, enabling it to be transmitted through the channel as symbols. At the receiver, inverse operations are performed to reconstruct the transmitted information. Communication systems typically pursue two optimization goals. Firstly, the reliability metric aims to minimize the discrepancy between the estimated sequence at the receiver and the transmitted sequence at the transmitter. In technical communication, this is evaluated using metrics like the BER, while in semantic communication, it considers semantic differences. Secondly, the efficiency metric aims to minimize the length of transmitted symbols. These objectives often conflict with each other. Shannon's separation theorem states that in technical communication systems, these objectives correspond to channel encoding and source encoding, respectively, and are optimized independently.

This study analyze{s} the problem of semantic communication from the perspective of optimal decoding, focusing on the process where an information sequence \linebreak  $X = (x_1,x_2,\cdots,x_n)$, with $x_i \in \mathbb{X}$ is transmitted. $\mathbb{X}$ represents the set of transmitted information symbols. Initially, $X$ is encoded and modulated at the transmitter using the function described in Equation (\ref{eq_semantic_encoder}), resulting in $S=(s_1,s_2,\cdots,s_t)$, where $s_i \in \mathbb{S}$ and $\mathbb{S}$ denotes the set of transmitted symbols. Subsequently, after transmission through the channel, the receiver acquires the sequence $O=(o_1,...,o_t)$. The receiver task is to estimate $\hat{S}$ from $O$, and $\hat{S}$ is then decoded into $\hat{X}$ using a semantic decoding function specified by Equation  (\ref{eq_semantic_decoder}). The optimization objectives of the communication system are twofold: Firstly, to  {minimize} information error{s} by aligning the probability distribution of $\hat{S}$ as close as possible to $S$. Here, the function $\varphi$ is bijective and reversible, ensuring that recovering $S$ is equivalent to recovering $X$. Secondly, the system aims to minimize the length $t$ of the sequence $S$, thereby improving the efficiency.
\begin{equation}
  \label{eq_semantic_encoder}
  S = \varphi  (X),
\end{equation}
\begin{equation}
  \label{eq_semantic_decoder}
  \hat{X} = \varphi^{-1} (\hat{S}),
\end{equation}

The 
 relationship between $o_t$ and $s_t$ is shown in Equation (\ref{eq1}):
\begin{equation}
  \label{eq1}
  o_t = h_t \otimes s_t + n_t \,
\end{equation}
where $h_t$ and $n_t$ denote the channel  {impulse} response (CSI) and noise at time $t$, and $\otimes$ signifies the convolution operator. The process is a typical Markov process, and the optimal decoding and demodulating process {involves solving the following:}
\begin{equation}
  \label{eq2}
  \hat{S} = \mathop{\arg\max}\limits_{(s_i \in \mathbb{S})}P(s_1,s_2,\cdots,s_t | o_1,o_2,\cdots,o_t),
\end{equation}

In fact, $P(S|O)$ is often unavailable, hence Bayesian equation in Equation (\ref{eq3}) is commonly employed to modify it:
\begin{equation}
  \label{eq3}
  P(S|O)= \frac{P(S,O)}{P(O)} = \frac{P(O|S)P(S)}{P(O)},
\end{equation}
where $P(O|S)$ represents the channel conditional transition probability. The design of the communication system assumes that the channel is memoryless, so the following formula  {is valid}:
\begin{equation}
  \label{eq5}
  P(o_1,o_2,\cdots,o_t|s_1,s_2,\cdots,s_t)=\prod_{i=1}^{t}P(o_i|s_i),
\end{equation}

 {Therefore, the optimization goal of the optimal decoding system is as follows:}
\begin{align}
  \label{eq_6}
  &\mathop{\arg\max}\limits_{(s_i \in \mathbb{S})} P(s_1, s_2, \cdots, s_t \mid o_1, o_2, \cdots o_t) 
  =& \mathop{\arg\max}\limits_{(s_i \in \mathbb{S})}\frac{P(s_1,s_2,\cdots,s_t) \prod_{i=1}^t P(o_i \mid s_i)}{P(o_1, o_2, \cdots, o_t)},
\end{align}

 {In Equation (\ref{eq_6}), $P(s_1,s_2,\cdots,s_T)$ represents the probability of the transmitted symbol sequence occurring within the entire set of transmitted symbols $\mathbb{S}$. For a specific received sequence, $P(o_1, o_2, \cdots, o_T)$ is a fixed value and cannot be optimized. Therefore, the final optimization objective is set by Equation  (\ref{eq_youhua2}).}\vspace{-8pt}
\begin{align}
  \label{eq_youhua2}
  &\mathop{\arg\max}\limits_{(s_i \in \mathbb{S})} P(s_1, s_2, \cdots, s_t \mid o_1, o_2, \cdots o_t) 
  =& \mathop{\arg\max}\limits_{(s_i \in \mathbb{S})}P(s_1,s_2,\cdots,s_t) \prod_{i=1}^t P(o_i \mid s_i)
\end{align}

The problem now revolves around identifying a transmission system where the receiver can access both $P(s_1, s_2, \ldots, s_t)$ and $P(o_i \mid s_i)$. The length of $t$ determines the coding rate. If such a system exists, it would be optimal  {in view of} the current coding rate. The two components of this requirement will be analyzed separately in the subsequent sections.

First, $P(o_i \mid s_i)$ represents the probability distribution of the received signal at the receiver through the channel. For an AWGN channel, this distribution is given by \linebreak Equation~(\ref{eq_10}):
\begin{equation}
  \label{eq_10}
  P\left(o_i \mid s_i\right)=\frac{1}{\sqrt{2 \pi} \sigma} e^{\left(-\frac{(s_i-o_i)^2}{2 \sigma^2}\right)},
  \end{equation}
where $\sigma^2$ denotes the noise power. Considering fading channels beyond AWGN channels, if $h_i$ can be accurately estimated at time $i$, then based on Equation (\ref{eq1}), $P\left(o_i \mid h_i \otimes s_i\right)$ follows a normal distribution, as expressed in Equation (\ref{eq_11}):
\begin{equation}
  \label{eq_11}
  P\left(o_i \mid h_i \otimes s_i\right)=\frac{1}{\sqrt{2 \pi} \sigma} e^{\left(-\frac{(o_i- s_i \otimes h_i)^2}{2 \sigma^2}\right)},
\end{equation}

\textls[-35]{When $h_i$ remains constant, the fading channel effectively behaves like an \mbox{AWGN~channel.}}

The `$\otimes$' in Equation (\ref{eq_11}) corresponds to channel equalization in classical technical communication systems, where $s_i$ represents the constellation sequences corresponding to the transmitted semantic symbol. In digital communication, the estimated $h_i$ is utilized to equalize $o_i$ by  {multiplying} the conjugate transposed matrix of $o_i$ and $h_i$. However, we  {use} $h_i$ to convolve $s_i$ to the advantage of the design of our subsequent system.

Thus, our objective transforms into identifying a function $\varphi$ that maps the sequence $X$ to the sequence $S$, while accurately modeling the probability of $S$. This probability model should be consistent and computable at both transmitter and receiver ends, where the length of $S$ reflects the coding rate at the transmitter. In technical communication systems, determining the probability distribution of $S$ is challenging, often assuming an equiprobable transmission of information bits. LLMs process text into numerical token sequences and predict the probabilities of subsequent tokens in natural language, closely resembling our problem. These models can effectively model $P(S)$. In subsequent sections, we detail the use of LLMs for the probabilistic modeling of text. Assuming that $P(S)$ is accessible, we continue discussing the function $\varphi$.

The goal of $\varphi$ is to minimize the length $t$ of $S$, thereby maximizing data transmission rates, ensuring that $P(s_1, s_2, \cdots, s_t)$ reaches its maximum:\vspace{-6pt}
\begin{equation}
  \max P(s_1, s_2, \cdots, s_t) = \prod_{i=1}^t P(s_i),  \nonumber
\end{equation}
\begin{equation}
  \label{eq_st}
  \text{subject to} \quad P(s_i) = P(s_j) \quad \forall \ 1 \leq i, j \leq t. 
\end{equation}

When transmitted symbols are independent and equiprobable, the mapping process from $X$ to $S$ approaches the entropy limit in compression. This scenario mirrors source coding achieving the entropy limit without channel coding in technical communication~\citep{weaver1953recent}. The optimal decoding strategy involves comparing the Euclidean distance between the received signal and the transmission constellation. Bit errors are irreparable without prior information; once an error occurs at the receiver, the information becomes unrecoverable. However, technical communication systems typically exhibit some correlation{s} among elements within $S$, enabling the receiver to correct erroneous symbols using prior knowledge. This correlation is analogous to the use of parity bits in channel coding within technical communication systems. Channel coding fundamentally entails computing parity bits from information bits using a corresponding generator matrix, thereby facilitating error recovery through receiver-side correlation. Viterbi decoding principles maximize the probability of information bits given the posterior distribution of received symbols. Hence, from both technical and semantic perspectives, an ideal communication system should possess the following characteristics:
\begin{itemize}
  \item  $S$ should be as short as possible, but cannot reach the entropy bound. The process from $X$ to $S$ is an encoding process that retains a certain amount of information redundancy.
  \item The information redundancy inherent in $S$ can be effectively exploited, allowing the design of algorithms to implement Equation (\ref{eq_youhua2}).
\end{itemize}

 {DeepSC \citep{xie2021deep} employs a Transformer-based architecture for joint semantic encoding and decoding. The encoder maps input text $X$ to a fixed-length semantic vector $S \in \mathbb{R}^d$, while the decoder reconstructs $\hat{X}$ from noisy channel outputs. Its training objective is to minimize semantic loss (e.g., sentence similarity) rather than bit-level errors, enabling robustness in low-SNR regimes but limiting error-free transmission at high SNRs.}

\subsection{System Model}
In this subsection, the application of LLMs for modeling the problems outlined in the previous subsection is expounded. LLMs undergo an unsupervised pre-training process aimed at predicting the probability of the next token in a training set, given preceding text. Tokens,  {which represent} encoded forms of human language beyond individual characters, are derived from extensive natural language data and  {undergo the phase of} tokenizer training, akin to a data compression process \citep{sennrich2015neural}.  {Frequently used} methods such as Byte Pair Encoding (BPE) and WordPiece \citep{wu2016google} serve as forms of data compression, aligning with the function $\varphi$ discussed earlier. The embedding layer after tokenization further represents semantic information from this encoding. 

Consider a scenario where both the transmitter and receiver share identical background knowledge from a common knowledge base. If all possible sequences of transmitted information can be enumerated, then $P(S)$ becomes computable, resolving the problem  {stated} in the foregoing subsection. The objective of LLM training is to predict token probabilities {after the} 
tokenization of text, while dataset construction aims to comprehensively cover all human languages. Assuming text information, Alice and Bob, needing communication, should possess the same knowledge base ($X$ distribution equivalence) for effective language interaction. Assuming a comprehensive collection of their language usage into dataset $\mathbb{D}$, and given adequate computational resources for both parties, algorithms like BPE or WordPiece tokenize $\mathbb{D}$’s text. The goal is  {to train} a tokenizer with a vocabulary size $V$. If $m$ bits represent a token, then $2^m \geq V$. Tokenization’s compression effect---where UTF-8 encodes sequence $X$ of length $n$ into sequence $W$ of length $t$---should satisfy the~expression
\begin{equation}
  \label{eq_st11}
  m*t  \leq   8*n,
  \end{equation}

Assuming \(V \) approaches infinity and the number of training iterations tends to infinity, tokenization achieves maximum data compression. For a dataset \(D \), this tokenization results in each token in the vocabulary becoming approximately equiprobable. Consequently, the encoding of transmitted information reaches its shortest form. If such tokenization is applied to train an LLM, the model may fail to converge, as the output \(P(S) \) remains constant. Therefore, by controlling the size of \(V \), one can regulate the degree of compression through tokenization, thereby managing the redundancy of information carried by transmitted bits. This effectively controls the coding rate of source--channel encoding \citep{rust2020good}.

Next, tokens are fed into LLMs for unsupervised pre-training. The model output can be approximated as
\begin{equation}
  \label{eq_9}
  LLM_{output} \approx P(w_i \mid w_{i-1}, w_{i-2}, \cdots, w_{max(1,i-N)}),
\end{equation}
where $N$ denotes the context length of the LLM. As LLMs evolve, $N$ tends to increase. Given the contextual dependencies inherent in natural language sequences, the tokens $W=(w_i, w_{i-1},  w_{i-2}, \cdots,  w_{max(1, i-N)})$ are not independent. Thus, the equation addressing the causal system is\vspace{-6pt}
\begin{equation}
  \label{eq4}
  P(W) =  \prod_{i=1}^{t} \! P(w_i | w_{i-1},  w_{i-2}, \cdots,  w_{max(1, i-N)}),
\end{equation}

 {The semantic knowledge base is established through the unsupervised pre-training phase of LLMs by Equation (\ref{eq4}), which learns the joint probability distribution $P(W)$ of token sequences from massive text corpora. }

The loss function for unsupervised pre-training of LLMs is
\begin{equation}
  \label{eq_llmloss}
  \mathcal{L} = -\sum_{i=1}^{t} \log P(w_i \mid w_{i-1}, w_{i-2}, \ldots, w_{max(i-N,1)}),
\end{equation}

It is observed that Equation (\ref{eq4}) and the loss function Equation (\ref{eq_llmloss}) are equivalent.  {This distribution serves two critical functions: (1) it guides the beam search decoder by ranking candidate sequences based on linguistic plausibility, and (2) it enables semantic error correction by assigning near-zero probabilities to implausible sequences. The knowledge base is static after pre-training and requires no fine-tuning for deployment.} By designing a high-dimensional constellation diagram for $W$ and modulating it, $W$ can becomes $S$. It should be noted that because the vocabulary size is generally large, it is difficult to represent it with one modulation symbol. Therefore, multiple modulation symbols are needed to represent a token, which we call a high-dimensional constellation diagram. Since $W$ and $S$ are in one-to-one correspondence, they share identical probability distributions. To train a semantic system for optimal decoding, the loss function is formulated as follows:\vspace{-6pt}
\begin{align}
  \label{eq_semanticloss}
  \mathcal{L_s} = -\sum_{i=1}^{t} \log P(w_i \mid w_{i-1}, w_{i-2}, \ldots, w_{max(i-N,1)}) - \sum_{i=1}^{t} \log P(o_i \mid w_i),
\end{align}

The two components of Equation (\ref{eq_semanticloss}) are evidently independent, enabling separate training of the system parts. The second component can be expressed as follows:\vspace{-6pt}
\begin{align}
  \label{eq_semanticloss2}
  - \sum_{i=1}^{t} \log P(o_i \mid w_i) = - \sum_{i=1}^{t} \log P(o_i \mid s_i),
\end{align}

Equation (\ref{eq_semanticloss2}) represents the channel condition transition probability. Once the channel and modulation scheme are determined, this equation can be specified as Equation (\ref{eq_11}). For modulation schemes akin to those in classical communication (e.g., QAM or QPSK), Equation (\ref{eq_semanticloss2}) can be directly applied without additional training.

Therefore, language sequences can be semantically encoded through the tokenization and training of LLMs. Probabilistic modeling is subsequently performed on these encoded sequences. Achieving a zero loss in LLM training suggests optimal decoding within this transmission system, where loss indicates deviation from optimal decoding. Notably, through the training of tokenization and LLMs, an optimal transmission system can be attained across various coding rates, obviating the need for further LLM fine-tuning. This training process aligns seamlessly with the typical training methodologies of LLMs, requiring no adjustments for semantic communication.

\begin{figure}[H]

  \includegraphics[width=5.4in]{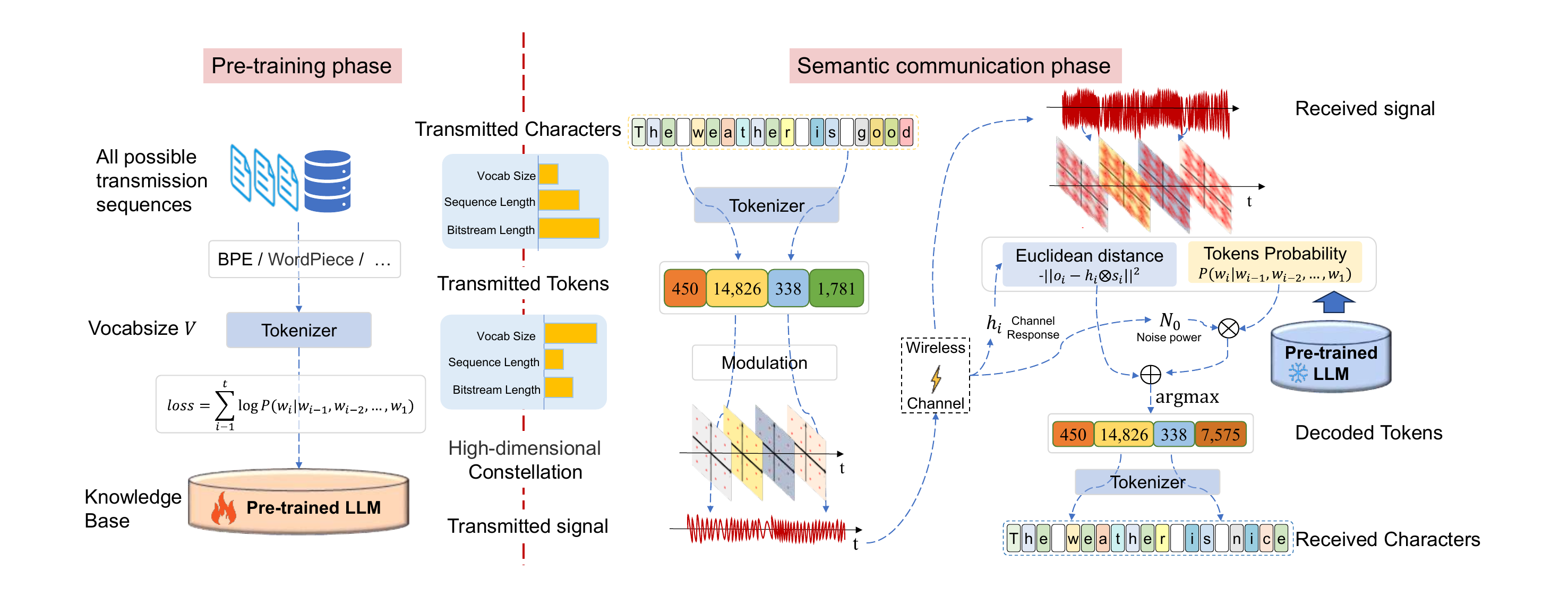}
  \caption{The 
 structure of LLM-SC. }
  \label{fig_system}
  \end{figure}

The system structure depicted in Figure \ref{fig_system} illustrates the pre-training process. We constructed the dataset by collecting all possible  {transmitted} contents from both the sender and receiver. A tokenizer with a vocabulary size of $V$ was trained using algorithms such as BPE and WordPiece. This tokenizer converts the text into tokens, which are  {later} used to train an LLM. The loss function for the LLM is defined in Equation (\ref{eq_llmloss}). Upon completion of the LLM training, the model serves as a knowledge base for semantic communication by providing $P(W)$ to supply shared knowledge to the receiver. At the receiving end, the LLM plays a crucial role in the decoder by offering priors for the transmitted sequence. The method for achieving optimal decoding will be discussed in subsequent sections. The LLM-SC has the following characteristics:

\begin{itemize}
  \item The system assumes that $D$  {comprises} all possible content that can be transmitted between the transmitter and receiver. While achieving this in {real life} poses challenges, current training corpora for LLMs strive to encompass all languages used by humans. Furthermore, this assumption is becoming increasingly feasible as LLM technology evolves~rapidly.
  \item The system prioritizes decoding sentences with higher probabilities of occurrence in the real world. Sequences for which $P(s_1,s_2,\ldots,s_T) = 0$ cannot be transmitted, as the receiver would be unable to decode such sequences where the probability is zero. Consequently, the system cannot transmit sentences that are impossible in the real world, as these sentences convey no meaningful information. Therefore, the system integrates both technical and semantic information.
\end{itemize}

\subsection{Decoding Algorithm of LLM-SC}

After training, a pre-trained LLM can compute the probability in Equation (\ref{eq4}) of a sent sequence $S=(s_1, s_2, \cdots, s_t)$ given a received sequence $O=(o_1, o_2, \cdots, o_t)$, representing the likelihood of the sequence $S$ transforming into $O$ after transmission through a channel. This decoding process can be applied using any pre-trained LLM, and in this subsection, we delve into the specific decoding algorithm.

 {According to Equation (\ref{eq_youhua2}), assuming the transmitted sequence length does not exceed the maximum context length of the LLM, the optimal decoding strategy involves}\vspace{-6pt}
\begin{align}
  \label{eq_8}
  \hat{W} &= \underset{w_i \in V}{\operatorname{argmax}} \prod_{i=1}^t P\left(w_i \mid w_{i-1}, w_{i-2}, \cdots, w_1\right) P\left(o_i \mid w_i\right)    \nonumber  \\
  &= \underset{w_i \in V}{\operatorname{argmax}} \prod_{i=1}^t P\left(w_i \mid w_{i-1}, w_{i-2}, \cdots, w_1\right)  P\left(o_i \mid s_i\right) \nonumber \\
  &= \underset{w_i \in V}{\operatorname{argmax}} \sum_{i=1}^t [\ln(P\left(w_i \mid w_{i-1}, w_{i-2}, \cdots, w_1\right)) + \ln(P\left(o_i \mid s_i\right))],
\end{align}

 {The first term in Equation (\ref{eq_8}) reflects the prior probability distribution of transmitted tokens, while the second term quantifies the Euclidean distance between received symbols and constellation points. The direct exploration of all possible sequences to select the one with the highest probability, known as maximum likelihood decoding, poses computational challenges with complexity $O(V^t)$, where $V$ is the number of possible states (sets of transmit symbols) and $t$ is the sequence length, typically imposing engineering constraints.}\vspace{-6pt}
  \begin{align}
    \label{eq_decoding}
    \hat{W} &= \underset{w_i \in V}{\operatorname{argmax}} \prod_{i=1}^t P\left(w_i \mid w_{i-1}, w_{i-2}, \cdots, w_1\right) P\left(o_i \mid w_i\right)    \nonumber  \\
    &= \underset{w_i \in V}{\operatorname{argmax}} \sum_{i=1}^t [  N_0 \! \ln(P\left(w_i  \mid  w_{i-1},  w_{i-2},  \cdots,  w_1\right))  -  \|o_i  -  h_i  \otimes  s_i\|^2],   
  \end{align}
where $N_0$ denotes the noise power spectral density (PSD), typically $N_0 = 2\sigma^2$. Here, $w_i$ represents the $i$-th transmitted token that undergoes modulation into $s_i$ and transmission through the channel. Equation (\ref{eq_decoding}) represents a typical hidden Markov model prediction problem, commonly addressed using dynamic programming algorithms to find the path with the maximum probability.  {The Viterbi algorithm provides optimal sequence decoding for hidden Markov models by dynamically tracking maximal-probability paths. However, its \mbox{$O(V^2 \times t)$} complexity is prohibitive for large vocabularies (\mbox{$V\!=\!$ 32,000}).}  To mitigate complexity, beam search, a heuristic algorithm prevalent in NLP, is adapted. Beam search sets up a search tree using breadth-first search, sacrificing optimality for efficiency. In beam search, the beam width \(K \) governs its operation. At each timestep, it retains the top \(K \) sequences by score, expanding each to generate \(K \times V \) candidates. From these, the next \(K \) best sequences are selected to continue expansion.

 {Beam search transitions into the Viterbi algorithm when \mbox{\(K = V \)}, and into a greedy decoding algorithm when \mbox{\(K = 1 \)}. It strikes a balance between the optimality of Viterbi and the efficiency of greedy decoding, making it suitable for scenarios where minor errors are acceptable given the expansive search space.}

Our decoding approach  {is modified} using the scoring function described by \linebreak  \mbox{Equation~(\ref{eq_12})}. Specifically, a max heap of size $K$ is maintained to store high-scoring sequences.  {Then}, $K$ sequences  {are used} to predict the next symbol and these are added to the heap. Then, we extract the current top $K$ for the next iteration:\vspace{-6pt}
\begin{equation}
  \label{eq_12}
  socre \! = \! N_0 \ln P(w_i \! \! \mid \! \! w_{i-1}, \! w_{i-2}, \! \cdots, \!  w_1) \!  -  \! \|o_i-h_i \! \otimes \! s\|^2,
\end{equation}

Beam search decoding with a computational complexity of \(O(K \times V \times t) \) facilitates parallelized computation across \(K \) sequences. Algorithm \ref{alg:alg2} outlines the workflow, designed to demodulate text efficiently. By adjusting \(K \), computational efficiency against decoding accuracy  {can be balanced}. Beam search offers a feasible, though suboptimal, approach leveraging the capabilities of LLMs. The specific decoding process is illustrated in Figure \ref{fig_system}.

     {Algorithm \ref{alg:alg2} details beam search decoding, where}
\begin{itemize}
  \item  Beam size 
 (\mbox{$K$}): Controls the trade-off between accuracy (higher \mbox{$K$}) and complexity;
  \item  Target tokens 
 (\mbox{$t$}): Predefined based on transmitted sequence length;
  \item  Noise PSD 
 (\mbox{$N_0$}): Scales the semantic prior term in Equation (\ref{eq_12}).
\end{itemize}

 {The output $W$ is the highest-scoring token sequence after $t$ iterations.}
 
   \begin{algorithm}[H]
    \caption{Beam search for text decoding 
 }
    \label{alg:alg2}
    \begin{algorithmic}
    \STATE 
    \STATE \textbf{Input: }$O$: Received symbols from wireless channel
    \STATE \hspace{1.0cm}$K$: beam size
    \STATE \hspace{1.0cm}$t$: the number of target decoding tokens
    \STATE \hspace{1.0cm}$N_0$: PSD of noise
    \STATE \textbf{Output: }$W:$ Decoding tokens
    \begin{enumerate}
    \STATE \hspace{0.3cm}$B_0 \leftarrow \{<0,>\}  $
    \STATE \hspace{0.3cm}\textbf{for} $i \in \{0,…,t-1\}:$
    \STATE \hspace{0.3cm}\hspace{0.5cm}$B \leftarrow \infty  $
    \STATE \hspace{0.3cm}\hspace{0.5cm}\textbf{for} $<sorce,\textbf{y}> \in B_i-1$
    \STATE \hspace{0.3cm}\hspace{1.0cm}\textbf{for} $s \in \mathcal{S}$
    \STATE \hspace{0.3cm}\hspace{1.5cm}$sorce \leftarrow N_0ln(P(s|\textbf{y}))-||o_i- h_i \! \otimes s||^2$
    \STATE \hspace{0.3cm}\hspace{1.0cm}$B.add(<sorce,\textbf{y}\circ s>)$
    \STATE \hspace{0.3cm}$B_t \leftarrow B.top(K)$
    \STATE \hspace{0.3cm}\textbf{return} $B.max()$
    \end{enumerate}
    \end{algorithmic}
    \label{alg2}
    \end{algorithm}

\subsection{Performance Metrics}
Performance metrics are essential for evaluating proposed methods in communication systems. In end-to-end communication, BER is commonly adopted as the training target by both transmitters and receivers, yet it often overlooks broader communication goals. BER may not accurately reflect the performance of semantic communication systems. Consequently, novel metrics such as Bilingual Evaluation Understudy (BLEU) and Word Error Rate (WER) have been proposed, focusing on word-level similarity between transmitter and receiver outputs. However, these metrics do not fully capture the similarity between entire sentences. To address this gap, metrics utilizing pre-trained models like BERT have emerged for evaluating semantic similarity. This paper selects evaluation metrics that encompass both traditional technical communication systems and emerging semantic communication paradigms, providing a comprehensive assessment of the proposed~method.

\textit(1) BLEU: BLEU is a metric commonly used to evaluate the quality of machine-generated text against one or more reference texts. Originally developed for machine translation, it has been adapted in semantic communication systems where assessing the quality of generated outputs is crucial. BLEU calculates a score based on the precision of $n$-grams (continuous sequences of $n$ items, typically words) between the generated output and the reference texts. It quantifies how closely the generated text matches the reference texts in terms of these $n$-grams, providing a numerical assessment of similarity and fluency. For a transmitted sentence $s$ of length $l_s$ and its decoded counterpart $\hat{s}$ with length $l_{\hat{s}}$, the BLEU score is calculated as follows:\vspace{-6pt}
\begin{equation}
  \label{eq_bleu}
  \log \text{BLEU} = \min (1-\frac{l_{\hat{s}}}{l_s},0) + \sum_{n=1}^{N} u_n\log p_n,
\end{equation}
where $u_n$ are weights assigned to $n$-grams, and $p_n$ denotes the $n$-gram score:
\begin{equation}
  \label{eq_bleu_pn}
  p_n = \frac{\sum _k \min (C_k(\hat{s}),C_k(s))}{\sum _k \min(C_k(\hat{s}))},
\end{equation}

Here, $C_k(\cdot)$ represents the frequency count function for the $k$-th elements in $n$-grams. BLEU captures contextual relationships to some extent but primarily evaluates superficial similarity of $n$-grams without considering semantic equivalence. Consequently, BLEU may assign a lower score even when the generated text is semantically aligned with the reference, due to differences in expressions or the use of synonyms.

(2) Sentence similarity: For addressing the issue of polysemous words, Xie {et al.} introduce a novel evaluation metric for sentence similarity \citep{xie2021deep}. Sentence similarity assesses the semantic equivalence between two sentences using a pre-trained model. Such models, exemplified by BERT \citep{devlin2018bert}, are natural language processing models trained extensively on diverse corpora. The semantic similarity between sentences \(s \) and \(\hat{s} \), both from the sender and receiver perspectives, is computed as follows:\vspace{-6pt}
\begin{equation}
  \label{eq_sentence_similarity}
  \text{match}(s,\hat{s}) = \frac{\mathbf{B_{\Phi }(s)} \mathbf{B_{\Phi }(\hat{s})^T}}{ \left\lVert \mathbf{B_{\Phi }(s)} \right\rVert  \left\lVert\mathbf{B_{\Phi }(\hat{s})} \right\rVert},
\end{equation}
where \(\mathbf{B}_{\Phi} \), based on BERT, represents a highly parameterized pre-trained model designed for extracting semantic information from text.

(3) BER: BER is a widely adopted metric for assessing the performance of communication systems, quantifying the likelihood of correctly receiving a transmitted symbol. A low BER signifies the system's capability to accurately recover transmitted symbols. However, bit-level errors often inadequately reflect the performance of semantic transmission systems. For instance, occasional bit errors per word may yield a negligible BER, yet render the received text unintelligible due to the absence of correctly received words. Conversely, texts with notably high BER can remain understandable if crucial semantic elements are correctly deciphered. Leveraging LLM-SC, which achieves error-free transmission and facilitates communication compatible across technical and semantic levels, BER serves as a pertinent evaluation metric.
 
(4) Token Error Rate (TER): The token acts as the fundamental unit for transmission and demodulation in LLM-SC. In addition to BER, errors in tokens reflect the challenge of accurately comprehending a sentence. For instance, a low BER combined with errors uniformly distributed across \(t \) tokens results in a high TER, indicating persistent difficulty in sentence comprehension by the receiver. Conversely, errors concentrated within a specific token, despite a potentially high BER, do not impair semantic understanding, thereby maintaining system effectiveness. This metric bears resemblance to the WER discussed in~\citep{farsad2018deep}. While each token can correspond to a word, they encompass a broader spectrum of entities and constitute the basic input units for LLMs.

\section{Numerical Results}
Simulations have been conducted to assess the capability of LLM-SC, focusing on measuring feasibility rather than extensive benchmarking. From the discussions above, it  {is} concluded that using existing LLMs allows for the verification of the feasibility of the proposed method, while the vocabulary size \(V \) remains fixed. Adjusting the modulation order provides a means to control communication efficiency. In this section, we present the existing LLMs that we employed without fine-tuning or additional training  to verify the feasibility of LLM-SC. We present a comprehensive comparison of existing semantic communication schemes among multiple evaluation metrics and discuss factors influencing their performance. Notably, this method accounts for the technical communication context, enabling restoration of original characters and facilitating comparison with traditional communication systems in terms of BER, a challenging metric to compare all semantic communication systems.

The dataset consists of English text extracted from the Europarl proceedings \citep{koehn2005europarl}, widely utilized in semantic communication tasks. Due to computational constraints, we conducted preprocessing to eliminate extraneous characters and filter out sentences that were excessively short or long.  {The LLM utilized in our simulations was Vicuna (v1.5), fine-tuned from LLaMA. Vicuna-7B v1.5 is chosen for its balance of efficiency and performance. Fine-tuned from LLaMA on conversational data, it achieves 90\% of ChatGPT's quality on MT-Bench \citep{zheng2023judging}. Its 4,096-token context window supports long-sequence decoding, and the 32,000-token vocabulary covers diverse linguistic constructs in the Europarl corpus. In fact, we can use any existing LLM to conduct experiments.} 

 {AWGN and Rayleigh fading channels were selected for their fundamental representation of real-world scenarios. The AWGN channel models ubiquitous thermal noise, while Rayleigh fading captures multipath effects in non-line-of-sight environments (e.g., urban areas). These models are standardized benchmarks in 3GPP specifications \citep{maxwell20185g}.} Table \ref{tab:experiment_set} outlines the key simulation settings.
\begin{table}[H]
\small
  \caption{Simulation settings. \label{tab:experiment_set}}
  \begin{tabularx}{\textwidth}{CL}
    \toprule
    \textbf{Parameters 
}& \textbf{Value 
 }\\
    \midrule
    LLM Model  & Vicuna 7b v1.5 \\
    \midrule
    Dataset    & European Parliament Proceeding\\
    \midrule
    CPU &  16 vCPU AMD EPYC 9654 96-Core Processor \\
    \midrule
    GPU      &  NVIDIA RTX 4090   \\
    \midrule
    Average number of characters of sent text (100,000 samples) &  348.87   \\
    \midrule
    Average number of tokens in sent text &   78.93   \\
    \midrule
    SNR (dB)  &  2--20  \\
    \midrule
    Temperature   &  1.0 \\
    \midrule
    Beam size   & 1--30  \\
    \midrule
    Modulation scheme   & 8-QAM/16-QAM  \\
    \midrule
    Vocabulary size & 32,000 \\
    \midrule
    The number of bits per token   & 15, 16 
  \\
    \midrule
    Channel model  & AWGN/Rayleigh \\
    \bottomrule
  \end{tabularx}
\end{table}

We conducted simulations using an NVIDIA RTX 4090 for performance evaluation. Specifically, 16QAM and 8-QAM modulations were adopted for comparisons. The choice of 16QAM facilitate{d} comparisons with traditional communication systems, while 8-QAM  {was} selected for comparison against the findings in the literature \citep{xie2021deep}. Our simulations encompassed AWGN and Rayleigh fading channel models, and include multiple sets of comparative analyses to assess the influence of beam size.

\subsection{The Performance of BLEU and Sentence Similarity}
To evaluate the performance of LLM-SC in comparison with existing popular methods in semantic communication systems, we utilized BLEU and sentence similarity as our evaluation metrics. Unlike the modulation method previously discussed, the methods presented in \citep{xie2021deep} employ various modulation schemes. To ensure a fair comparison with the work in \citep{xie2021deep}, we used bits per transmitted symbol (bps) as the metric for coding efficiency, which quantifies the number of information bits conveyed by each modulation symbol. According to \citep{xie2021deep}, their coding efficiency is approximately $1.07$ bps. Hence, for LLM-SC to achieve comparable coding efficiency, we employed a 15 bits per token encoding strategy coupled with an 8-QAM modulation scheme, wherein each token is represented by five modulation symbols in a high-dimensional constellation. This approach ensures that the coding efficiency of LLM-SC aligns with that of the DeepSC method proposed in~\citep{xie2021deep}, thereby maintaining equivalence in the number of transmitted characters for a given number of~symbols.

\begin{figure}[H]
  \subfloat[AWGN channel. ]{\includegraphics[width=1\textwidth]{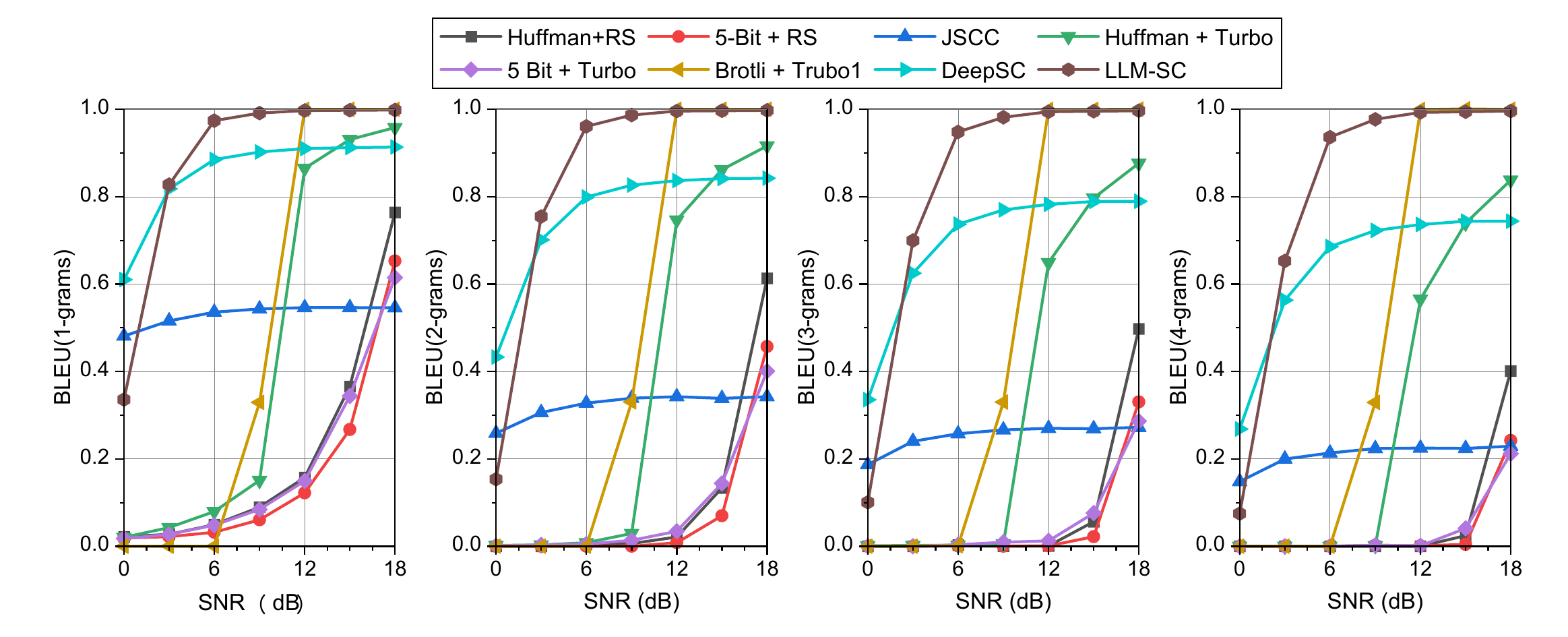}}
\\
  \subfloat[Rayleigh fading channel. ]{\includegraphics[width=1\textwidth]{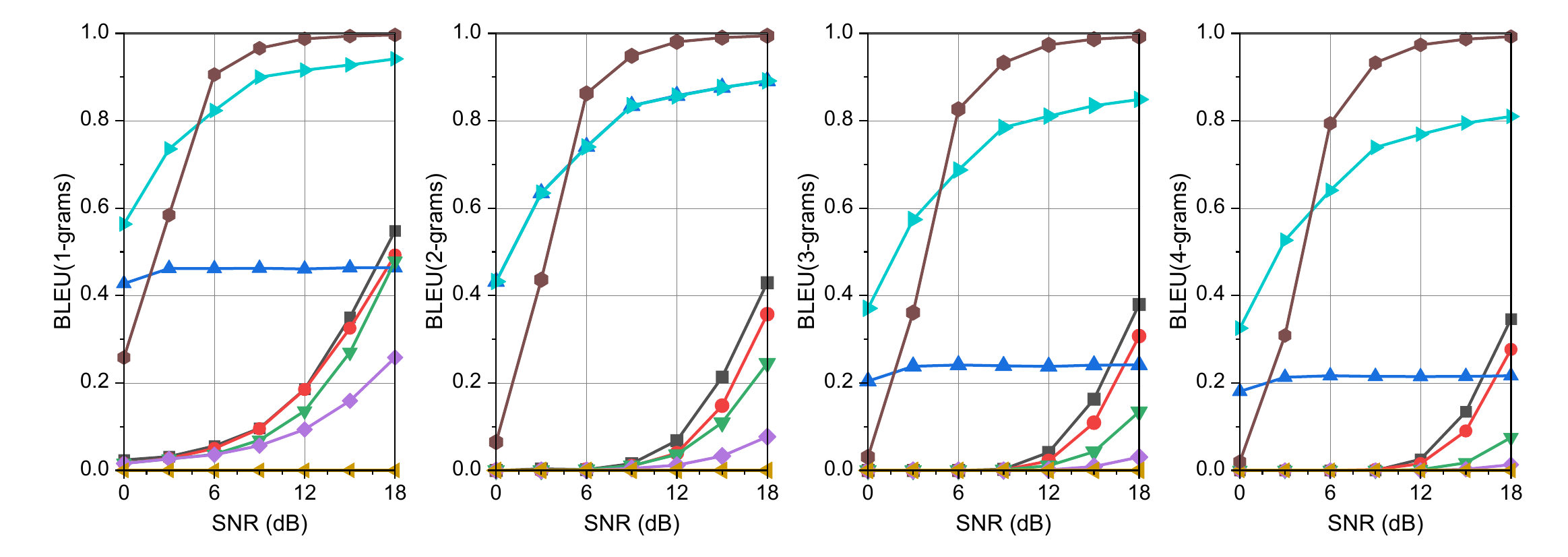}}
  \caption{BLEU scores versus the SNR for the same bps, with Huffman coding with RS (30,42) in \linebreak  64-QAM, 5-bit coding with RS (42,54) in 64-QAM, Huffman coding with Turbo coding in 64-QAM, 5-bit coding with Turbo coding in 128-QAM, Brotli coding with Turbo coding in 8-QAM,~the DNN-based JSCC~\citep{park2020end} trained over AWGN channels and Rayleigh fading channels, DeepSC trained over the AWGN channels and Rayleigh fading channels \citep{xie2021deep}, and finally, our proposed LLMS-SC.}
  \label{fig_result_bleu}
\end{figure}

Figure \ref{fig_result_bleu} illustrates the relationship between the BLEU score and the SNR for identical bps over AWGN and Rayleigh fading channels. The comparison includes traditional techniques such as Huffman coding with RS (30,42) in 8-QAM, 5-bit coding with RS (42,54) in 64-QAM, Huffman coding with Turbo coding in 64-QAM, 5-bit coding with Turbo coding in 128-QAM, Brotli coding with Turbo coding in 8-QAM, and the DNN-based JSCC trained over AWGN channels and Rayleigh fading channels, as reported in \citep{xie2021deep}, alongside the DeepSC approach, which employs trained modulation maps.

It can be observed in Figure \ref{fig_result_bleu}a that AI-based methods generally outperform traditional source--channel separation coding methods. Specifically, at SNR levels below 3 dB, the DeepSC model demonstrates the  {best} performance in terms of BLEU score, indicating its superior adaptability to channel noise under low-SNR conditions. Conversely, for SNR levels above 3 dB, the LLM-SC model exhibits superior BLEU performance across all n-grams, significantly surpassing traditional coding schemes. This suggests that, under high-SNR conditions, the LLM-SC model is more effective in recovering semantic information. Notably, the DeepSC model fails to achieve completely error-free transmission even at very-high-SNR levels, implying that DeepSC cannot ensure the reliable transmission of all semantic information. In contrast, the LLM-SC model achieves a BLEU score of 1 at high-SNR levels, with near-zero errors at SNRs above 10 dB. Additionally, the BLEU performance of LLM-SC improves rapidly with increasing SNRs. When comparing BLEU scores across multiple n-grams, it is evident that the degradation of the BLEU score with increasing n in n-grams is slowest for the LLM-SC method. This is reflected in the diminishing advantage of DeepSC over LLM-SC with increasing n at low-SNR levels, whereas at high-SNR levels, the advantage of LLM-SC over DeepSC  {strengthens} with larger n-grams. This indicates that the LLM-SC method is more favorable for semantic coherence, prioritizing the decoding of semantically continuous tokens, which aligns with the characteristics of LLMs.
\vspace{-7pt}

Figure \ref{fig_result_bleu}b shows the performance of LLM-SC under Rayleigh fading channels. Traditional encoding and decoding schemes experience significant performance degradation in Rayleigh channels, with the highest BLEU score not exceeding 0.6 at an SNR of 18 dB. In contrast, DeepSC shows better adaptability to Rayleigh fading channels, indicating that neural network-based semantic encoding and decoding systems can effectively learn channel characteristics. However, DeepSC still fails to achieve error-free transmission, even when BLEU scores are high. The LLM-SC model demonstrates performance under Rayleigh fading channels similar to that in AWGN channels, with no noticeable degradation, consistent with the conclusion of Equation (\ref{eq_11}). Furthermore, LLM-SC achieves error-free transmission at high-SNR levels and outperforms all compared models. When comparing multiple n-gram scores, the advantage of LLM-SC over other methods  {heightens} with higher n-grams at high-SNR levels.

\begin{figure}[H]
  \subfloat[AWGN channel. ]{\includegraphics[width=0.45\textwidth]{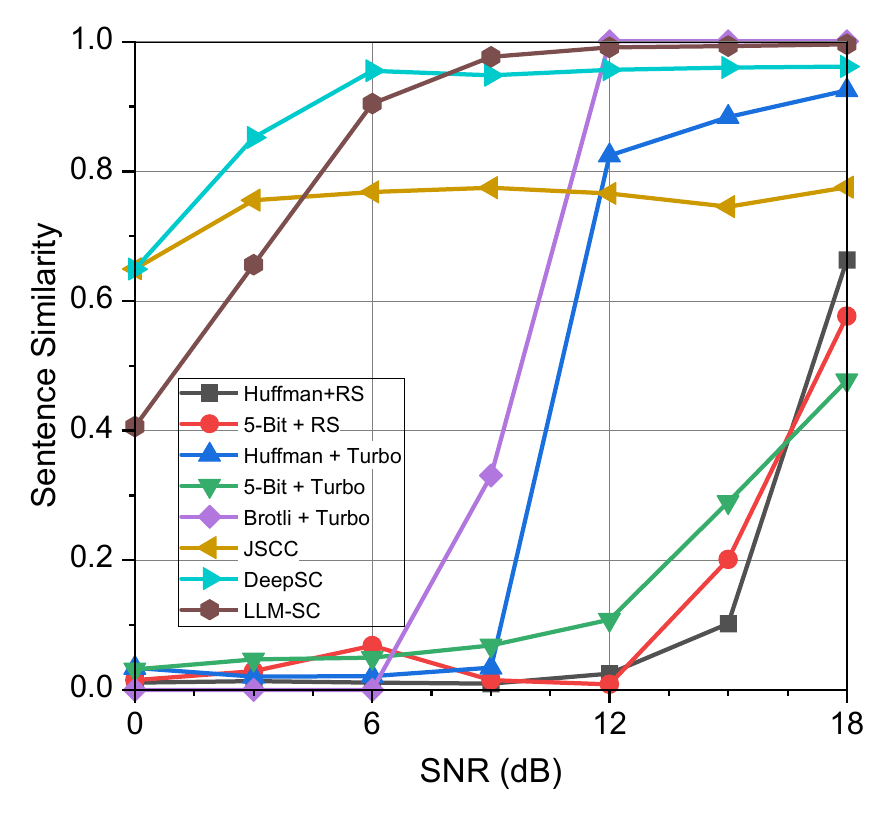}}
  \hfill
  \subfloat[Rayleigh fading channel. ]{\includegraphics[width=0.45\textwidth]{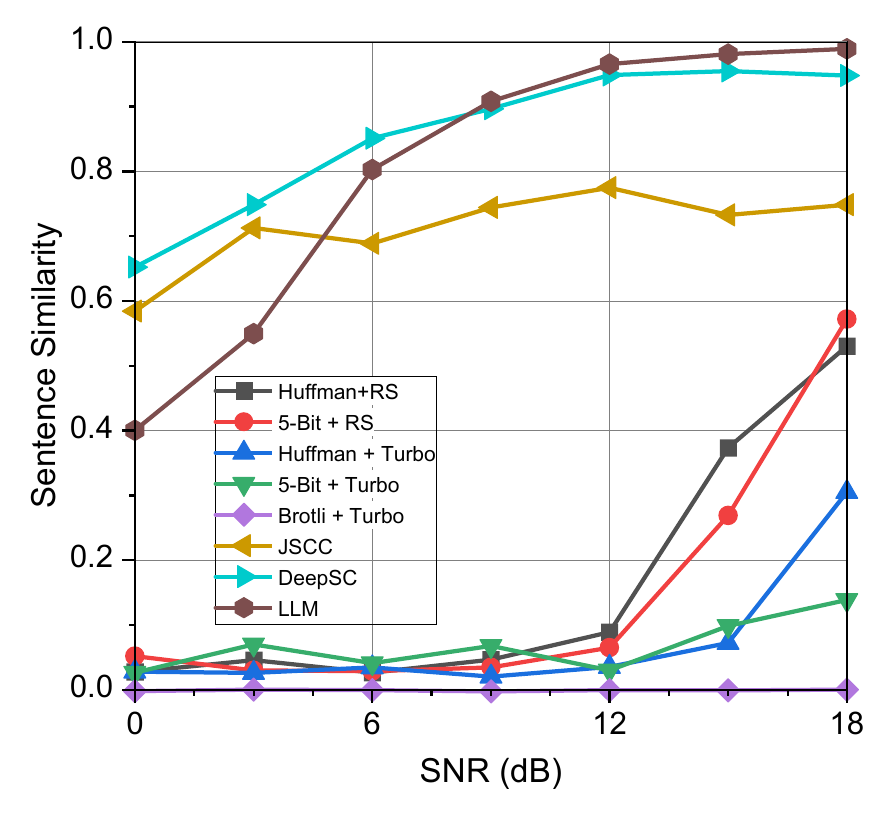}}
  \caption{Sentence similarity versus SNRs for the same number of bits per  transmitted symbol, with Huffman coding with RS (30,42) in 64-QAM, 5-bit coding with RS (42,54) in 64-QAM, Huffman coding with Turbo coding in 64-QAM, 5-bit coding with Turbo coding in 128-QAM, Brotli coding with Turbo coding in 8-QAM,~the DNN-based JSCC trained over AWGN channels and Rayleigh fading channels~\citep{park2020end}, DeepSC trained over the AWGN channels and Rayleigh fading channels \citep{xie2021deep}, and finally, our proposed LLM-SC.}
  \label{fig_result_sentence_similarity}
\end{figure}

Figure \ref{fig_result_sentence_similarity} presents a comparison of sentence similarity metrics. In Figure \ref{fig_result_sentence_similarity}a, all methods exhibit the same increasing trend in sentence similarity as the SNR increases. Traditional methods such as Huffman with Turbo coding achieve a BLEU (1-gram)  score of 0.2 under 9~dB, yet the sentence similarity approaches 0. This indicates that even if some words are received, the sentence cannot be properly understood. Machine learning-based methods like DeepSC and LLM-SC show consistency between sentence similarity and BLEU scores. At low SNRs, DeepSC has higher similarity than LLM-SC, but DeepSC has an upper bound across the entire SNR range and cannot achieve error-free transmission similar to the BLEU score. In contrast, LLM-SC demonstrates better competitiveness at high SNR, capable of transmitting complete semantic information. In Figure \ref{fig_result_sentence_similarity}b, under the Rayleigh fading channel, all methods experience significant degradation compared to the AWGN channel, indicating that channel fading has a substantial impact on semantic transmission. In this scenario, LLM-SC still maintains a high advantage at high SNRs, achieving near-error-free transmission, but performs slightly worse than DeepSC and JSCC at low SNRs due to the impact of channel fading.\vspace{-14pt}

\subsection{The Performance of BER and TER}
To validate the effectiveness and reliability of the LLM-SC, we also conducted a comparison with traditional technical communication systems. Based on the observations from the previous subsection, the performance of the LLM-SC under AWGN and Rayleigh fading channels is comparable;  therefore, the simulations in this subsection were conducted only under AWGN channels.

Unlike BLEU and sentence similarity, which are often used for evaluating semantic communication, both LLM-SC and classical algorithms utilize classical modulation schemes such as QAM and QPSK. Therefore, we compared the BER and TER under the same modulation scheme, allowing us to assess system performance under identical modulation and joint coding efficiency conditions. In this simulation, we employed the commonly used 16-QAM modulation scheme for a fair comparison. To ensure a fair comparison, we first calculated the equivalent joint source--channel coding rate of LLM-SC using the metric bits per symbol (bpc), which represents the number of bits used for each character transmitted through the channel. In technical communication systems, this value is typically the product of the source coding rate and the channel coding rate. Hence, we compute the average equivalent joint source--channel coding rate for LLM-SC. Based on simulation data presented in Table \ref{tab:experiment_set}, the equivalent joint coding rate of LLM-SC is approximately $R \approx \frac{78.93 \times 16}{348.87} \approx 3.62$ (bpc). The factor of $16$ arises from the use of a vocabulary size of 32,000, necessitating at least 15 bits for representation, along with the employment of 16-QAM modulation. Consequently, in a high-dimensional constellation, at least four modulation symbols are required to represent a token. Since 16-QAM carries 4 bits per modulation symbol, 16 bits are needed per token to be encoded. The approximation ($\approx$) indicates that this rate is derived from a large number of sentences, and may vary across different sentences. This detailed analysis ensures a rigorous comparison of the LLM-SC system against classical communication systems under the same modulation conditions, providing insights into their relative performance in terms of BER and TER.

To benchmark against traditional technical communication systems, we calculate the coding rate of commonly used source coding methods, such as Huffman coding, zlib, and arithmetic coding (AC), also expressed in bpc, as shown in Table \ref{tab:compression}.\vspace{-3pt}

\begin{table}[H]
  \caption{Coding rate of some text compression algorithms. \label{tab:compression}}
  \begin{tabularx}{\textwidth}{CC}
    \toprule
  \textbf{Coding  Algorithm} 
 & \textbf{Coding Rate (bpc)} 
 \\
    \midrule
    Huffman  & 3.736 \\
    \midrule
    Zlib & 4.904 \\
    \midrule
    AC + MultiPPM & 5.76 \\
    \midrule
    \bf{LLM-SC 
}  & \bf{3.62 
} \\
    \bottomrule
  \end{tabularx}
\end{table}\vspace{-3pt}
As shown in Table \ref{tab:compression}, the joint coding efficiency of LLM-SC surpasses that of conventional text compression algorithms. Notably, LLM-SC maintains semantic-level redundancy, which enhances error correction capabilities at the receiver. In contrast, for technical communication systems, achieving the same bpc precludes the incorporation of additional channel coding schemes, as these would increase the bpc. Nevertheless, even without considering channel coding, the bpc of technical systems remains higher to that of LLM-SC. Therefore, for subsequent comparisons, channel coding schemes will be excluded for technical communication systems. 

We evaluated the demodulation performance by transmitting English text samples from the dataset, comparing LLM-SC in 16-QAM with UTF-8 encoding (hard demodulation) in 16-QAM. The metrics used for evaluation  {were} BER and TER, as tokens represent the fundamental unit of transmission and reception. The beam size utilized in the simulation was 15.

\begin{figure}[H]
  \includegraphics[width=4.5in]{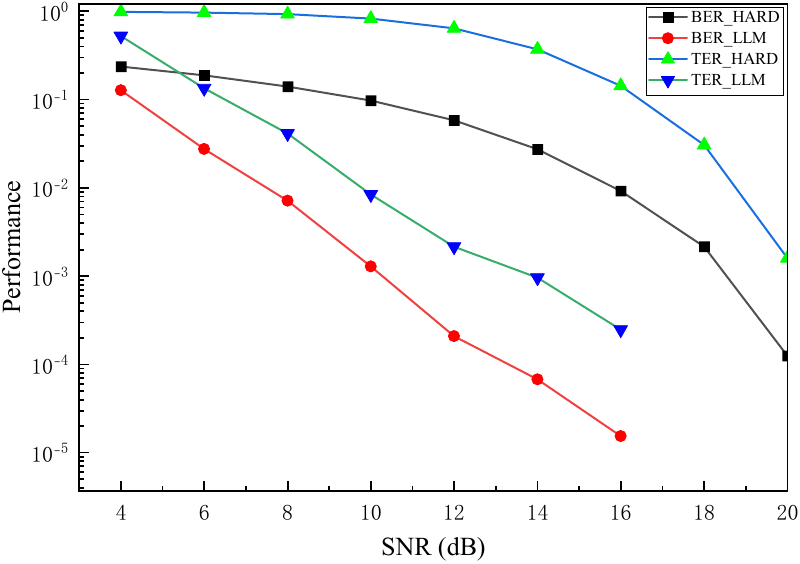}
  \caption {The 
 BER and TER performance of LLM-SC and hard demodulation.}
  \label{fig_result}
\end{figure}

The results, presented in Figure \ref{fig_result}, demonstrate that LLM-SC significantly outperforms hard demodulation. For BER performance, LLM-SC exhibits strong competitiveness across the entire SNR curve. Notably, as SNR rises, the BER of LLM-SC declines rapidly, while the decline for the hard demodulation method is much slower. At the common $10^{-3}$ BER threshold, LLM-SC achieves a coding gain of approximately 8 dB. Furthermore, no bit errors  {are} observed in the simulation for SNR values exceeding 16 dB, indicating that LLM-SC can achieve error-free bit transmission in technical communication systems.

Regarding TER, hard demodulation struggles at SNR values below 14 dB, whereas LLM-SC reaches a $10^{-3}$ TER at 14 dB, successfully decoding the majority of words. The improvement in TER is consistent across all SNR levels, as LLM-SC leverages contextual information. Similarly, TER decreases faster with increasing SNR, and no token errors  {are found} for SNR values above 16 dB, indicating that LLM-SC can achieve error-free token transmission in technical communications.

In summary, the superior coding efficiency and the preservation of semantic redundancy for error correction underscore the advantages of LLM-SC over traditional methods, positioning LLM-SC as a more effective alternative for semantic communication tasks.
\vspace{-5pt}

\subsection{Effect of Beam Size}
\begin{figure}[H]
\hspace{-8pt}  \includegraphics[width=4.0in]{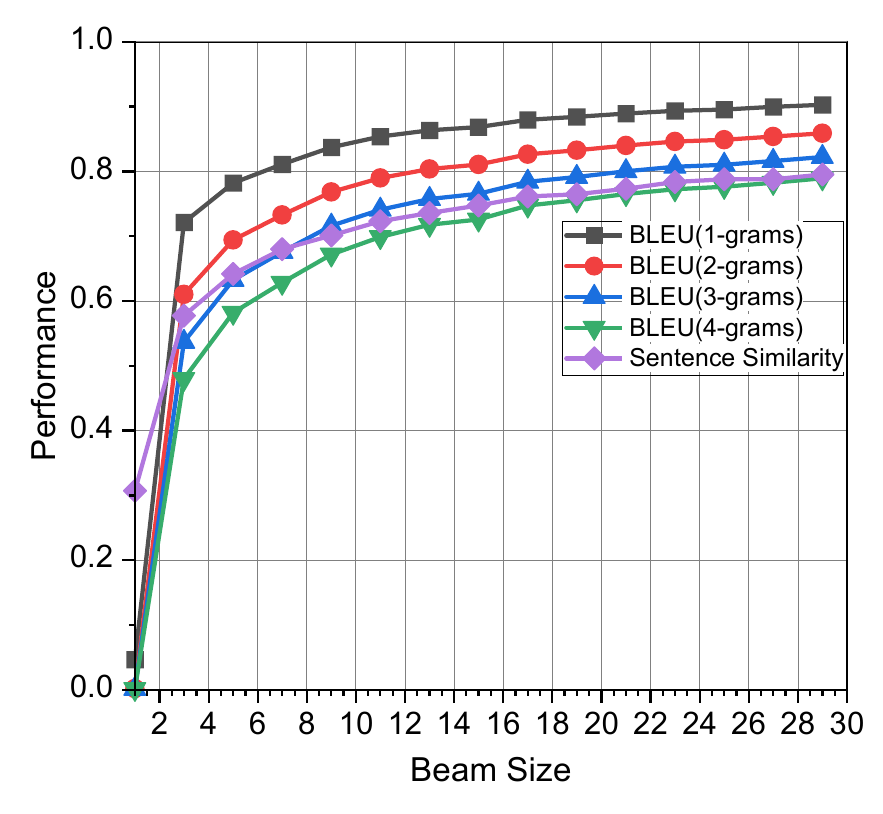}
  \caption{Performance of LLM-SC with  different beam search widths.}
  \label{fig_beam}
\end{figure}

Decoding performance heavily depends on the beam search width \(K \). We conducted simulations comparing various \(K \) values, averaging BLEU and sentence similarity performance at each SNR, as shown in Figure~\ref{fig_beam}. Narrow beams (\(K = 1 \)) performed the poorest, akin to greedy search, resulting in the lowest BLEU and sentence similarity scores.  {Three key patterns emerged from the averaged results:}

\begin{enumerate}
    \item \textls[-25]{Significant Gain from Small $K$:
 {Increasing $K$ from 1 to 5 yields substantial~improvements: }}
    \begin{itemize}
        \item  {BLEU-1 increases by $\Delta 0.73$ (from $\sim$0.05 to $\sim$0.78);}
        \item  {Sentence similarity increases by $\Delta 0.27$ (from $\sim$0.31 to $\sim$0.58).}
    \end{itemize}
     {This reveals beam search's critical advantage over greedy decoding ($K=1$).}
    
    \item Diminishing Returns Beyond $K=10$: 
{Further increasing $K$ from 10 to 30 provides minimal gains:}
    \begin{itemize}
        \item  {BLEU-1 improves by only \mbox{$\Delta 0.07$} (from {$\sim$0.85} to {$\sim$0.92});}
        \item  {Sentence similarity improves by \mbox{$\Delta 0.07$} (from {$\sim$0.71} to {$\sim$0.78}).}
    \end{itemize}
     {The relative gain drops below 10\% for $K>10$, indicating rapidly diminishing returns.}
    
    \item Optimal Operating Point: 
 {$K=10$ achieves >90\% of the maximum observed performance for both metrics. This suggests $K_{\text{opt}} \approx 10$ provides the best accuracy-complexity trade-off for practical deployment.}
\end{enumerate}

 {These findings accord with the theoretical expectation: LLM token distributions are typically concentrated on a few high-probability candidates. Beyond the top-$K$ paths (where $K$ matches the typical number of plausible continuations), additional exploration yields minimal benefits.}

In summary, a moderate \(K \) strikes a balance between efficiency and accuracy. Values around 10 optimize both performance and complexity by concentrating the search on the most probable sequences informed by language model statistics. This approach leverages the inherent knowledge of the language model to prune implausible decoding.

\vspace{-8pt}

\section{Discussion}
\subsection{Case Analysis and Interpretations}
Table \ref{tab:llm_demodulation} shows a comparison of the demodulating capabilities of the LLM-SC and DeepSC over Rayleigh fading channels. At a 6 dB SNR, DeepSC can demodulate some words,  {realizing} a 1-gram BLEU score of 0.54. However, it is challenging to extract useful information from the received content, while its BER is as high as 0.39. In contrast, LLM-SC achieves a 1-gram BLEU score of 0.94 at 6dB SNR, allowing us to obtain most of the intended semantics from the transmitted sentence, with a BER of only 0.026. At 12 dB SNR, DeepSC reaches a BLEU score of 0.72, enabling partial understanding of the transmitted information, though accurately grasping the semantics remains difficult, with a BER of 0.32. Comparatively, LLM-SC achieves error-free transmission at 12 dB SNR, allowing complete recovery of the transmitted sentence, with optimal BLEU and BER performance. Additionally, at 3 dB SNR, LLM-SC still achieves a BLEU score of 0.86, enabling us to understand most of the critical information from the received data. Of course, the sampling selected a commonly used English sentence with strong context relevance. The results of the average performance are shown in Figure \ref{fig_result_bleu}.

A deeper analysis reveals that LLM-SC concentrates errors on a limited number of tokens, allowing for the demodulation of the majority of tokens and thus enabling a high level of information understanding. In contrast, DeepSC prioritizes improving BLEU scores by considering entire sentences. While it correctly demodulates many words, resulting in a higher BLEU score, semantic comprehension remains imperfect. This underscores the limitations of using BLEU as a metric for semantic communication. Importantly, LLM-SC achieves a harmonious integration of technical and semantic communication, with metrics consistently evaluating both aspects. Unlike DeepSC, which excels semantically in BLEU but struggles technically with BER, and unlike traditional technical communication, which excels BER but falls short semantically, LLM-SC effectively balances both dimensions.

In essence, leveraging mutual information between symbols and meaning{s}, LLM-SC reliably transmits semantic content despite noise. Rather than solely maximizing symbolic fidelity, the system preserves information at higher linguistic levels—the essence of effective~communication.

\begin{table}[H]
  \caption{The sample sentences between different methods over Rayleigh channels. \label{tab:llm_demodulation}}

\begin{adjustwidth}{-\extralength}{0cm}
\centering
  \begin{tabularx}{\fulllength}{cCcc}
    \toprule
    \bf{Method} & \bf{Content} & \bf{BLEU (1-Gram)} & \bf{BER}  \\
    \midrule
    \bf{transmitted sentence 
} & life is just a series of trying to make up your mind about what you want to do and then doing it. & - & - \\
    \midrule
    \bf{LLM-SC (12 dB) 
} & life is just a series of trying to make up your mind about what you want to do and then doing it. & 1 & 0 \\
    \midrule
    \bf{LLM-SC (6 dB) 
} & life is just a series of steps to make up your mind about what you want to do and then doing it.  & 0.94 & 0.026  \\
    \midrule
    \bf{LLM-SC (3 dB) 
} & hardly ever just a game of trying to make up your mind about what you want to do and then doing it.& 0.86 & 0.031 \\
    \midrule
    \bf{DeepSC (12 dB) 
} & {funding is just a number of clearly to make up your note about what you want to do that and then at}  &0.72 & 0.32 \\
    \midrule
    \bf{DeepSC (6 dB) 
} & {secondly is just a m of having to make up your speaking about what you to have to do now so how at at}  & 0.54 & 0.39\\
    \bottomrule
  \end{tabularx}
\end{adjustwidth}
\end{table}

\subsection{Complexity Considerations}
The computational complexity of LLM-SC and DeepSC is compared in Table \ref{tab:llm_complexity} in terms of the average demodulating runtime per character. It is evident that the runtime of LLM-SC significantly exceeds that of DeepSC, primarily due to the vast number of parameters in LLMs. The assumptions in this paper presume unlimited computational power at both the transmitter and receiver. However, LLM-SC introduces a novel paradigm for semantic communication systems. With the exponential growth in computational power, real-time demodulation by both transmitter and receiver becomes achievable, potentially mitigating the complexity concerns. Moreover, comparing the runtime of machine learning-based semantic communication methods on GPUs is inherently biased, as practical communication devices are unlikely to integrate high-performance GPUs at both ends.

\begin{table}[H]  
  \caption{An example of LLM demodulation. \label{tab:llm_complexity}}
  \begin{tabularx}{\textwidth}{CC}
    \toprule
    \bf{Method} & \bf{Time}    \\
    \midrule
    \bf{LLM-SC 
} & 9.2 ms/character  \\
    \midrule
    \bf{DeepSC 
} & 1.24 ms/character  \\
    \bottomrule
  \end{tabularx}
\end{table}

\subsection{Insights and Challenges}
 {Relative to classical DeepSC, LLM-SC manifests several distinguishing characteristics:}
\begin{itemize}
  \item Longer context length: Word encoding and decoding lengths can match LLM's context length, whereas DeepSC's maximum length is limited to 30. According to Shannon's information theory, longer code lengths theoretically enhance error correction~\mbox{capability.}
  \item DeepSC disregards distinctions such as uppercase vs. lowercase, punctuation, and special characters, simplifying semantic encoding. In contrast, LLMs consider the entire natural language vocabulary, enhancing adaptability and alignment with actual~usage.
  \item DeepSC's training on the European Parliament dataset restricts its adaptability to other corpora, while LLMs are typically trained on broad natural language corpora, potentially making them applicable across multiple languages.
  \item DeepSC outputs fixed-length symbols regardless of input length, whereas LLM-SC adapts symbol length based on input, potentially improving efficiency.
\end{itemize}

However, challenges remain in terms of computation constraints and real-time requirements before fully harnessing LLMs' potential in semantic communication. Advances in model architecture, accelerators, compression, and quantization methods can mitigate these challenges. Future avenues include benchmarking different model architectures, analyzing artifacts, enhancing robustness, and conducting comparative studies across diverse datasets and languages. Exploring joint optimizations with classical error-correcting codes also holds promise.

\section{Conclusions}
This paper introduces a novel LLM-SC framework for textual data within wireless communication systems. {{We propose} that leveraging the tokenizer of an LLM {serves as} an effective joint source{--}channel coder}. 
An optimal LLM-enabled decoding and demodulation method is derived to resiliently demodulate text by integrating the LLM's contextual understanding. It is  {deduced} that the pre-training of LLMs fundamentally constructs the encoding and decoding mechanisms for semantic communication, achievable without altering the original training process. Existing pre-trained models can be utilized for semantic encoding and decoding.

Extensive simulations demonstrate that LLM-SC outperforms traditional communication systems in terms of bit error rate (BER) at the technical communication level and outcompetes current machine learning-based methods in semantic-level metrics. Compared to technical communication algorithms, LLM-SC excels in extracting contextual relationships from text and leveraging the receiver LLM's contextual understanding for error correction, thereby achieving lower BER. In comparison to other semantic communication systems, LLM-SC tends to demodulate meaningful sentences, thus outperforming in BLEU and sentence similarity metrics.  {Furthermore,} the LLM acts as a semantic knowledge base in these systems, providing {both }sender and receiver with a probabilistic distribution of transmitted~sequences.

In conclusion, this pioneering study demonstrates initial feasibility and motivates further research in co-designing LLMs to advance intelligent communication systems. Beyond maximizing bit transmission, integrating higher-level semantics is crucial for unlocking future capabilities. This work makes a significant step toward LLM-empowered wireless systems focused on meaningful transmission.


\authorcontributions{Conceptualization, Z.W. and R.L.; methodology, Z.W.; software, L.Z.; validation, S.W.; formal analysis, F.L.; investigation, K.L.; resources, L.Z.; data curation, R.L.; \mbox{writing---original} draft preparation, Z.W.; writing---review and editing, K.L. and H.M.; visualization, R.L.; supervision, F.L.; project administration, Z.W. All authors have read and agreed to the published version of the manuscript.}

\funding{This 
 research was supported in part by the Natural Science Foundation of China under Grant No. U2441226.}

\institutionalreview{ Ethical review and approval were waived for this study
because it is an in vitro study, and there is no human and animal participation.  
}

\informedconsent{Not applicable.} 

\dataavailability{The source code and associated scripts (V1.0) for this study are openly available on GitHub 
 at 
\url{https://github.com/gujianhunwang/LLM\_com} accessed on 15 May 2025.} 

\conflictsofinterest{The authors declare no conflicts of interest.} 

\newpage
\abbreviations{Abbreviations}{
 {The following abbreviations are used in this manuscript:}
\\

\noindent 
\begin{tabular}{@{}ll}
    LLM & Large Language Model \\
    NLP  & Natural Language Processing \\
    BER    & Bit Error Ratio\\
    SNR &  Signal-to-Noise Ratio \\
    AI      &   Artificial Intelligence   \\
    JSCC &  Joint Source--Channel Coding   \\
    CSI &    Channel Impulse Response   \\
    PSD  &   Power Spectral Density  \\
    BLEU   &   Bilingual Evaluation Understudy \\
    WER   & Word Error Rate  \\
    TER  &  Token Error Rate \\
    bps & Bits Per Transmitted Symbol \\
    AC   & Arithmetic Coding  \\
    BERT & Bidirectional Encoder Representations from Transformers \\
\end{tabular}
}
\noindent



\begin{adjustwidth}{-\extralength}{-0cm}

\reftitle{References}




\PublishersNote{}
\end{adjustwidth}

\begin{thebibliography}{999}

\bibitem[Shannon(1948)]{shannon1948mathematical}
Shannon, C.E.
\newblock A mathematical theory of communication.
\newblock {\em  Bell Syst. Technol. J.} {\bf 1948}, {\em
27},~379--423. [\href{http://doi.org/10.1002/j.1538-7305.1948.tb01338.x}{CrossRef}]

\bibitem[Weaver(1953)]{weaver1953recent}
Weaver, W.
\newblock Recent contributions to the mathematical theory of communication.
\newblock In {\em ETC: A Review of General Semantics}; Institute of General Semantics: New York, NY, USA,  
{ 1953}; pp. 261--281.

\bibitem[Carnap et~al.(1952)Carnap, Bar-Hillel, et~al.]{carnap1952outline}
Carnap, R.; Bar-Hillel, Y.
\newblock \emph{An Outline of a Theory of Semantic Information};  Massachusetts
Institute of Technology:  Cambridge, MA, USA,  
{ 1952}.

\bibitem[Shao et~al.(2024)Shao, Cao, and G{\"u}nd{\"u}z]{shao2024theory}
Shao, Y.; Cao, Q.; G{\"u}nd{\"u}z, D.
\newblock A theory of semantic communication.
\newblock {\em IEEE Trans. Mob. Comput.} {\bf 2024}, \emph{23}, 12211--12228. [\href{http://dx.doi.org/10.1109/TMC.2024.3406375}{CrossRef}]


\bibitem[Xie et~al.(2021)Xie, Qin, Li, and Juang]{xie2021deep}
Xie, H.; Qin, Z.; Li, G.Y.; Juang, B.H.
\newblock Deep learning enabled semantic communication systems.
\newblock {\em IEEE Trans. Signal Process.} {\bf 2021}, {\em
69},~2663--2675. [\href{http://dx.doi.org/10.1109/TSP.2021.3071210}{CrossRef}]

\bibitem[Zhou et~al.(2021)Zhou, Li, Zhao, Peng, and Zhang]{zhou2021semantic}
Zhou, Q.; Li, R.; Zhao, Z.; Peng, C.; Zhang, H.
\newblock Semantic communication with adaptive universal transformer.
\newblock {\em IEEE Wirel. Commun. Lett.} {\bf 2021}, {\em
11},~453--457. [\href{http://dx.doi.org/10.1109/LWC.2021.3132067}{CrossRef}]

\bibitem[Xie et~al.(2021)Xie, Qin, and Li]{xie2021task}
Xie, H.; Qin, Z.; Li, G.Y.
\newblock Task-oriented multi-user semantic communications for VQA.
\newblock {\em IEEE Wirel. Commun. Lett.} {\bf 2021}, {\em
11},~553--557. [\href{http://dx.doi.org/10.1109/LWC.2021.3136045}{CrossRef}]

\bibitem[Weng et~al.(2021)Weng, Qin, and Li]{weng2021semantic}
Weng, Z.; Qin, Z.; Li, G.Y.
\newblock Semantic communications for speech signals.
\newblock In Proceedings of the ICC 2021-IEEE International Conference on
Communications, Montreal, QC, Canada, 14--23 June  2021; 
pp. 1--6.

\bibitem[Huang et~al.(2021)Huang, Tao, Gao, and Lu]{huang2021deep}
Huang, D.; Tao, X.; Gao, F.; Lu, J.
\newblock Deep learning-based image semantic coding for semantic
communications.
\newblock In Proceedings of the 2021 IEEE Global Communications Conference
(GLOBECOM),  Madrid, Spain, 7--11 December 2021;  
pp. 1--6.

\bibitem[Wang et~al.(2022)Wang, Dai, Liang, Niu, Si, Dong, Qin, and
Zhang]{wang2022wireless}
Wang, S.; Dai, J.; Liang, Z.; Niu, K.; Si, Z.; Dong, C.; Qin, X.; Zhang, P.
\newblock Wireless deep video semantic transmission.
\newblock {\em IEEE J. Sel. Areas Commun.} {\bf 2022},
{\em 41},~214--229. [\href{http://dx.doi.org/10.1109/JSAC.2022.3221977}{CrossRef}]

\bibitem[Chang et~al.(2024)Chang, Wang, Wang, Wu, Yang, Zhu, Chen, Yi, Wang,
Wang, et~al.]{chang2024survey}
Chang, Y.; Wang, X.; Wang, J.; Wu, Y.; Yang, L.; Zhu, K.; Chen, H.; Yi, X.;
Wang, C.; Wang, Y.;  et~al.
\newblock A survey on evaluation of large language models.
\newblock {\em ACM Trans. Intell. Syst. Technol.} {\bf
2024}, {\em 15},~1--45. [\href{http://dx.doi.org/10.1145/3641289}{CrossRef}]

\bibitem[Mohammed and Kora(2025)]{11015742}
Mohammed, A.; Kora, R.
\newblock A Comprehensive Overview and Analysis of Large Language Models:
Trends and Challenges.
\newblock {\em IEEE Access} {\bf 2025}, {\em 13},~95851--95875. [\href{http://dx.doi.org/10.1109/ACCESS.2025.3573955}{CrossRef}]

\bibitem[Guo et~al.(2023)Guo, Wang, Li, and Saeed]{guo2023semantic}
Guo, S.; Wang, Y.; Li, S.; Saeed, N.
\newblock Semantic Communications with Ordered Importance using ChatGPT.
\newblock {\em arXiv } {\bf 2023}, arXiv:2302.07142.

\bibitem[Shi et~al.(2018)Shi, Li, and Xie]{shi2018semantic}
Shi, G.; Li, Y.; Xie, X.
\newblock Semantic communications: Outcome of the intelligence era.
\newblock {\em Pattern Recognit. Artif. Intell.} {\bf 2018},
{\em 31},~91--99.



\bibitem[Azzouz et~al.(1996)Azzouz, Nandi, Azzouz, and
Nandi]{azzouz1996modulation}
Azzouz, E.E.; Nandi, A.K.; Azzouz, E.E.; Nandi, A.K.
\newblock Modulation recognition using artificial neural networks.
\newblock In {\em Automatic Modulation Recognition of Communication Signals}; Springer:  Berlin/Heidelberg, Germany, 
{
1996}; pp. 132--176.

\bibitem[Suetrong et~al.(2024)Suetrong, Taparugssanagorn, and
Promsuk]{suetrong2024enhanced}
Suetrong, N.; Taparugssanagorn, A.; Promsuk, N.
\newblock Enhanced Modulation Recognition Through Deep Transfer Learning in
Hybrid Graph Convolutional Networks.
\newblock {\em IEEE Access} {\bf 2024}, \emph{12},  54553--54566. [\href{http://dx.doi.org/10.1109/ACCESS.2024.3388490}{CrossRef}]


\bibitem[Jia et~al.(2023)Jia, Huang, Luo, and Wen]{jia2023lightweight}
Jia, Y.; Huang, Z.; Luo, K.; Wen, W.
\newblock Lightweight Joint Source-Channel Coding for Semantic Communications.
\newblock {\em IEEE Commun. Lett.} {\bf 2023}, \emph{12},  18447--18450. [\href{http://dx.doi.org/10.1109/LCOMM.2023.3329533}{CrossRef}]


\bibitem[Choi et~al.(2018)Choi, Tatwawadi, Weissman, and Ermon]{choi2018necst}
Choi, K.; Tatwawadi, K.; Weissman, T.; Ermon, S.
\newblock NECST: Neural joint source-channel coding. \emph{arXiv} {\bf 2018}, arXiv:1811.07557. 


\bibitem[Wang et~al.(2024)Wang, Guan, He, Hrovat, Liu, Zhong, Al-Dulaimi, and
Yu]{wang2024graph}
Wang, X.; Guan, K.; He, D.; Hrovat, A.; Liu, R.; Zhong, Z.; Al-Dulaimi, A.; Yu,
K.
\newblock Graph Neural Network enabled Propagation Graph Method for Channel
Modeling.
\newblock {\em IEEE Trans. Veh. Technol.} {\bf 2024}, \emph{73}, 12280--12289. [\href{http://dx.doi.org/10.1109/TVT.2024.3382650}{CrossRef}]


\bibitem[Yuan et~al.(2018)Yuan, Wu, Cheng, and Yang]{yuan2018deep}
Yuan, C.; Wu, C.; Cheng, D.; Yang, Y.
\newblock Deep learning in encoding and decoding of polar codes.
\newblock In Proceedings of the 2018 2nd International Conference on Data Mining, Communications and Information Technology (DMCIT 2018), 
Shanghai, China, \mbox{25--27 May 2018;}  
Volume  1060, p. 012021.

\bibitem[Soltani et~al.(2019)Soltani, Pourahmadi, Mirzaei, and
Sheikhzadeh]{soltani2019deep}
Soltani, M.; Pourahmadi, V.; Mirzaei, A.; Sheikhzadeh, H.
\newblock Deep learning-based channel estimation.
\newblock {\em IEEE Commun. Lett.} {\bf 2019}, {\em 23},~652--655. [\href{http://dx.doi.org/10.1109/LCOMM.2019.2898944}{CrossRef}]

\bibitem[Hekland et~al.(2009)Hekland, Floor, and Ramstad]{hekland2009shannon}
Hekland, F.; Floor, P.A.; Ramstad, T.A.
\newblock Shannon-kotel-nikov mappings in joint source-channel coding.
\newblock {\em IEEE Trans. Commun.} {\bf 2009}, {\em
57},~94--105. [\href{http://dx.doi.org/10.1109/TCOMM.2009.0901.070075}{CrossRef}]

\bibitem[Devlin et~al.(2018)Devlin, Chang, Lee, and Toutanova]{devlin2018bert}
Devlin, J.; Chang, M.W.; Lee, K.; Toutanova, K.
\newblock Bert: Pre-training of deep bidirectional transformers for language
understanding.
\newblock {\em arXiv } {\bf 2018}, arXiv:1810.04805.

\bibitem[Radford et~al.(2018)Radford, Narasimhan, Salimans, Sutskever,
et~al.]{radford2018improving}
Radford, A.; Narasimhan, K.; Salimans, T.; Sutskever, I.
\newblock Improving Language Understanding by Generative Pre-Training.
2018. 
\newblock Available online: \url{https://cdn.openai.com/research-covers/language-unsupervised/language\_understanding\_paper.pdf} ({accessed on 23 June 2025}).
%


\bibitem[Mikolov et~al.(2013)Mikolov, Chen, Corrado, and
Dean]{mikolov2013efficient}
Mikolov, T.; Chen, K.; Corrado, G.; Dean, J.
\newblock Efficient estimation of word representations in vector space.
\newblock {\em arXiv  } {\bf 2013}, arXiv:1301.3781.

\bibitem[Pennington et~al.(2014)Pennington, Socher, and
Manning]{pennington2014glove}
Pennington, J.; Socher, R.; Manning, C.D.
\newblock Glove: Global vectors for word representation.
\newblock In Proceedings of the 2014 Conference on Empirical
Methods in Natural Language Processing (EMNLP),  Doha, Qatar, 25--29 October 2014;  
pp. 1532--1543.

\bibitem[Berant et~al.(2013)Berant, Chou, Frostig, and
Liang]{berant2013semantic}
Berant, J.; Chou, A.; Frostig, R.; Liang, P.
\newblock Semantic parsing on freebase from question-answer pairs.
\newblock In  Proceedings of the 2013 Conference on Empirical
Methods in Natural Language Processing, Seattle, DC, USA, 8--21 October 2013;  
pp. 1533--1544.

\bibitem[Fellbaum(2010)]{fellbaum2010wordnet}
Fellbaum, C.
\newblock WordNet. In {\em Theory and Applications of Ontology: Computer
Applications}; Springer: Berlin/Heidelberg, Germany, 
2010; pp. 231--243.

\bibitem[Joulin et~al.(2016)Joulin, Grave, Bojanowski, and
Mikolov]{joulin2016bag}
Joulin, A.; Grave, E.; Bojanowski, P.; Mikolov, T.
\newblock Bag of tricks for efficient text classification.
\newblock {\em arXiv  } {\bf 2016}, arXiv:1607.01759.

\bibitem[Alhammadi et~al.(2024)Alhammadi, Shayea, El-Saleh, Azmi, Ismail,
Kouhalvandi, and Saad]{alhammadi2024artificial}
Alhammadi, A.; Shayea, I.; El-Saleh, A.A.; Azmi, M.H.; Ismail, Z.H.;
Kouhalvandi, L.; Saad, S.A.
\newblock Artificial Intelligence in 6G Wireless Networks: Opportunities,
Applications, and Challenges.
\newblock {\em Int. J. Intell. Syst.} {\bf 2024}, {\em
2024},~8845070. [\href{http://dx.doi.org/10.1155/2024/8845070}{CrossRef}]

\bibitem[Jiang et~al.(2023)Jiang, Peng, Dong, Wang, Yang, Pan, and
You]{jiang2023large}
Jiang, F.; Peng, Y.; Dong, L.; Wang, K.; Yang, K.; Pan, C.; You, X.
\newblock Large AI Model Empowered Multimodal Semantic Communications.
\newblock {\em arXiv  } {\bf 2023}, arXiv:2309.01249. [\href{http://dx.doi.org/10.1109/MCOM.001.2300575}{CrossRef}]

\bibitem[Jiang et~al.(2024)Jiang, Peng, Dong, Wang, Yang, Pan, and
You]{jiang2024large}
Jiang, F.; Peng, Y.; Dong, L.; Wang, K.; Yang, K.; Pan, C.; You, X.
\newblock Large AI model-based semantic communications.
\newblock {\em IEEE Wirel. Commun.} {\bf 2024}, {\em 31},~68--75. [\href{http://dx.doi.org/10.1109/MWC.001.2300346}{CrossRef}]

\bibitem[Shen et~al.(2024)Shen, Shao, Zhang, Lin, Pan, Li, Zhang, and
Letaief]{shen2024large}
Shen, Y.; Shao, J.; Zhang, X.; Lin, Z.; Pan, H.; Li, D.; Zhang, J.; Letaief,
K.B.
\newblock Large language models empowered autonomous edge AI for connected
intelligence.
\newblock {\em IEEE Commun. Mag.} {\bf 2024}, \emph{62}, 140--146. [\href{http://dx.doi.org/10.1109/MCOM.001.2300550}{CrossRef}]


\bibitem[Valmeekam et~al.(2023)Valmeekam, Narayanan, Kalathil, Chamberland, and
Shakkottai]{valmeekam2023llmzip}
Valmeekam, C.S.K.; Narayanan, K.; Kalathil, D.; Chamberland, J.F.; Shakkottai,
S.
\newblock LLMZip: Lossless Text Compression using Large Language Models.
\newblock {\em arXiv  } {\bf 2023}, arXiv:2306.04050.

\bibitem[Zhao et~al.(2024)Zhao, Yue, Hou, Cheng, and Huang]{zhao2024lamosc}
Zhao, Y.; Yue, Y.; Hou, S.; Cheng, B.; Huang, Y.
\newblock LaMoSC: Large Language Model-Driven Semantic Communication System for
Visual Transmission.
\newblock {\em IEEE Trans. Cogn. Commun. Netw.}
{\bf 2024}, \emph{10},  2005--2018. [\href{http://dx.doi.org/10.1109/TCCN.2024.3401712}{CrossRef}]


\bibitem[Chah(2018)]{chah2018ok}
Chah, N.
\newblock OK Google, What Is Your Ontology? Or: Exploring Freebase
Classification to Understand Google's Knowledge Graph.
\newblock {\em arXiv  } {\bf 2018}, arXiv:1805.03885.

\bibitem[Sennrich et~al.(2015)Sennrich, Haddow, and Birch]{sennrich2015neural}
Sennrich, R.; Haddow, B.; Birch, A.
\newblock Neural machine translation of rare words with subword units.
\newblock {\em arXiv  } {\bf 2015}, arXiv:1508.07909.

\bibitem[Wu et~al.(2016)Wu, Schuster, Chen, Le, Norouzi, Macherey, Krikun, Cao,
Gao, Macherey, et~al.]{wu2016google}
Wu, Y.; Schuster, M.; Chen, Z.; Le, Q.V.; Norouzi, M.; Macherey, W.; Krikun,
M.; Cao, Y.; Gao, Q.; Macherey, K.;  et~al.
\newblock Google's neural machine translation system: Bridging the gap between
human and machine translation.
\newblock {\em arXiv  } {\bf 2016}, arXiv:1609.08144.

\bibitem[Rust et~al.(2020)Rust, Pfeiffer, Vuli{\'c}, Ruder, and
Gurevych]{rust2020good}
Rust, P.; Pfeiffer, J.; Vuli{\'c}, I.; Ruder, S.; Gurevych, I.
\newblock How good is your tokenizer? on the monolingual performance of
multilingual language models.
\newblock {\em arXiv  } {\bf 2020}, arXiv:2012.15613.

\bibitem[Farsad et~al.(2018)Farsad, Rao, and Goldsmith]{farsad2018deep}
Farsad, N.; Rao, M.; Goldsmith, A.
\newblock Deep learning for joint source-channel coding of text.
\newblock In Proceedings of the 2018 IEEE International Conference on
Acoustics, Speech and Signal Processing (ICASSP), Calgary, AB, Canada, 15--20 April 2018;  
\mbox{pp.~2326--2330.}

\bibitem[Koehn(2005)]{koehn2005europarl}
Koehn, P.
\newblock Europarl: A parallel corpus for statistical machine translation.
\newblock In  Proceedings of the Machine Translation Summit X:
Papers, Phuket, Thailand, 13--15 September 2005;  
pp. 79--86.

\bibitem[Zheng et~al.(2023)Zheng, Chiang, Sheng, Zhuang, Wu, Zhuang, Lin, Li,
Li, Xing, et~al.]{zheng2023judging}
Zheng, L.; Chiang, W.L.; Sheng, Y.; Zhuang, S.; Wu, Z.; Zhuang, Y.; Lin, Z.;
Li, Z.; Li, D.; Xing, E.;  et~al.
\newblock Judging LLM-as-a-judge with MT-Bench and Chatbot Arena.
\newblock {\em arXiv  } {\bf 2023}, arXiv:2306.05685.

\bibitem[Maxwell(2018)]{maxwell20185g}
Maxwell, J.C.
\newblock 5G NR User Equipment (UE) radio transmission and reception; Part 1:
Range 1 standalone release 15 V. 15.2. 0 document 3GPP TS 38.101--1.
\newblock {\em  Treatise Electr. Magn.} {\bf 2018}, {\em
2},~68--73.

\bibitem[Park et~al.(2020)Park, Simeone, and Kang]{park2020end}
Park, S.; Simeone, O.; Kang, J.
\newblock End-to-end fast training of communication links without a channel
model via online meta-learning.
\newblock In Proceedings of the 2020 IEEE 21st International Workshop on Signal
Processing Advances in Wireless Communications (SPAWC), Virtual, 26--29 May 2020;  
pp.
1--5.

\end{thebibliography}
\end{document}